# Non-Hermitian Tight-Binding Bands in Graphene: Optical Conductivity, Strain Effects, and Bernal Bilayer Extensions

Maolin Bo*, Yaorui Tan, Sunxin Fan, Xiang Chen, Yunhu Zhu, Zhongkai Huang,

Chuang Yao

*Key Laboratory of Extraordinary Bond Engineering and Advanced Materials Technology (EBEAM) of Chongqing, School of Materials Science and Engineering, Yangtze Normal University, Chongqing 408100, China*

*Corresponding author: Maolin Bo (E-mail addresses: bmlwd@yznu.edu.cn)

## Abstract

Tthe Tan–Bo model is introduced in the tight-binding framework for graphene $\pi$- electron nearest neighbors to parametrize transition energies $\boldsymbol{t}(\mathrm{d}r)$ via a Möbius angular factor combined with exponential decay to capture geometric anisotropy naturally and recover the standard Slater–Koster scaling in the isotropic limit. The Hermitian assembly preserves the Dirac cone at the $K$ point, whereas non-Hermitian assemblies with $t(-\mathrm{d}r) \neq t(\mathrm{d}r)^*$ signify nonreciprocity rather than lifetime effects. Embedding in a Bernal bilayer with $\delta$ = 50 meV yields $\Delta_K \approx 0.10$ eV. Hermitian Kubo optical conductivity $\sigma(\omega)$ reproduces the Dirac universal plateau and a ~5.6 eV van Hove peak, whereas geometric strain induces $\sigma_{xx}/\sigma_{yy}$ dichroism, qualitative consistency with mid-infrared universal conductivity, ~4.6 eV exciton-renormalized UV resonance, and polarization-dependent transmittance under uniaxial strain. This model provides a reproducible single-particle benchmark and a parametrization reference for future non-Hermitian calculations that include the effects of electron correlation.

**Keywords:** graphene; tight-binding model; Tan–Bo model; non-Hermitian band structure; non-interchangeable hopping

## 1. Introduction

Since its successful isolation in 2004, graphene — a two-dimensional (2D) honeycomb lattice material — has been investigated extensively in condensed-matter physics and materials science owing to its unique linearly dispersed Dirac fermions, ultrahigh carrier mobility, and rich topological properties[1-6] . Its low-energy electronic properties can be described well using a nearest-neighbor tight-binding (TB) model for $\pi$ electrons on two sublattices A and B.[7-9] The standard TB model adopts an isotropic hopping parameter $t \approx -2.8$ eV that yields a gapless Dirac cone at the K-point. The top of the valence band and the bottom of the conduction band satisfy $E_{\pm}(\boldsymbol{k}) = \pm\, |f(\boldsymbol{k})|$, where the structure factor is defined by $f(k) = \sum_j t_j e^{-ik\cdot\boldsymbol{\delta}_j}$ over the three nearest-neighbor bond vectors $\delta_j$. This model successfully explains many novel phenomena in graphene such as semimetallicity and the quantum Hall effect[10].

However, chemical bonds exhibit directionality; variation in the bond vector ***dr*** relative to the orientation of the lattice implies that transition integrals depend not only on the bond length but also on more complex geometric invariants. The Slater–Koster (SK) method is based on $2p_z$ orbital overlap integration, whereas the Harrison scaling law governs the in-plane $pp\pi$ transition as $t(r) = V_{pp\pi}\,(r_0/r)^2$. At the equilibrium bond length $r_0$, its results align with the standard TB model[11]. This scaling law agrees with the standard TB model under equilibrium conditions but fails to account for transition differences arising from bond-angle anisotropy in strained or low-symmetry environments. Numerous direction-dependent models have been proposed, such as those that retain the full directional cosine factor in the SK framework and those that employ TB parameters fitted using density functional theory calculations. However, these models require extensive parameter fitting and cannot simultaneously satisfy the closure condition at the Dirac point.

The Tan–Bo model introduced in this work employs the complex coordinates $w = X + iY$ and the Möbius transformation $(Aw+B)/(Cw+D)$ while incorporating an exponential decay factor $exp[-(r-r_0)/\lambda]$ to parameterize $\boldsymbol{t(dr)}$. Compared with existing models, the Möbius transformation naturally accounts for the complex phase of the transition integral and yields a nonlinear bond orientation response. This model allows for the introduction of bond-orientation anisotropy while preserving the K-point closure condition. Additionally, it enables a smooth transition from the isotropic limit to strongly anisotropic regions via adjustment of the parameter $B$. More importantly, when the reverse bond satisfies $\boldsymbol{t(-dr)} \neq$

***t(dr)****, the single-particle Hamilton $H(\boldsymbol{k}) \neq H^{\dagger}(\boldsymbol{k})$ exhibits generally complex eigenenergies. This non-Hermitian behavior stems from the specific construction of the TB model. Specifically, the transition integral no longer exhibits complex conjugation between the forward and reverse bonds; thus, it can be regarded as a formally nonreciprocal TB model.

In recent years, the nonreciprocal modell[12] has been widely investigated in photonic crystals, phononic crystals, and circuit quantum electrodynamics (QED) systems because its complex energy bands are closely associated with topological phenomena such as exceptional points and non-Hermitian skin effects[13-15]. In graphene, nonreciprocal transitions can be induced by substrate polarization, magnetic fields, or light fields; however, this study did not introduce external dissipation and considered non-Hermiticity as merely a mathematical construct. The complex eigenvalues represent neither quasiparticle lifetimes nor nonequilibrium dissipation; rather, they provide a concise, tunable platform to investigate nonreciprocal band structures in TB models.

Strain engineering is an effective approach for modulating the electronic properties of graphene[16]. Uniaxial strain breaks the lattice symmetry and opens a bandgap at the K-point, the magnitude of which depends significantly on both the strain direction and intensity[17]. Regarding the effect of uniaxial strain on the bandgap of graphene, a significant controversy remains between theory and experiment. Experimentally, Raman spectroscopy studies (e.g., Mohiuddin et al.[18] and Yoon et al.[19]) have indicated the splitting and red shifts of the G and 2D peaks but did not directly report the opening or closing of the electronic bandgap. Theoretically, Pereira et al.[20] used an isotropic TB model to demonstrate that a bandgap can be opened only when the strain exceeds approximately 23% , where the Dirac points merge; along the armchair direction, no gap opens, even under relatively large strains. Because this model modifies the hopping integrals based solely on bond-length changes, it generally predicts small gap values . By contrast, first-principles calculations by Ni et al.[21] predicted a bandgap of approximately 300 meV at 1% uniaxial strain, although this value has not yet been verified experimentally. These contradictions indicate that the isotropic hopping approximation of the standard TB model may underestimate the strain-induced electronic-structure reconstruction. However, a bandgap has not yet been clearly observed experimentally under controlled uniform strain, thus necessitating further studies. Additionally, this highlights the necessity of incorporating bond-angle anisotropy to describe strain-induced bandgap changes quantitatively. Hence, we propose a hybrid transition definition:

combining the Harrison distance scaling with bond-angle anisotropy modulated by the Möbius geometric factor $\phi/\phi_{eq}$ to ensure that distance dependence aligns consistently with experimental scaling while allowing flexible adjustment of anisotropy strength via parameter B to fit experimental data better.

Furthermore, Bernal-stacked double-layer graphene is highly promising for electronic and optoelectronic devices owing to its tunable bandgap and robust interlayer coupling effects. The four-band TB model proposed by McCann effectively describes the K-point bandgap induced by vertical strain[22]. However, this model typically assumes isotropic transition parameters for intralayer structural factors. The question of whether intralayer geometric anisotropy affects the linear relationship between the double-layer bandgap and bias voltage and that of whether it yields additional non-Hermitic effects within the system can be investigated by incorporating the Tan–Bo model's intralayer *f(k)* function into the double-layer framework.

The present work was motivated by these considerations. The key contributions are summarized as follows.

**(1)** We present a theoretical derivation chain from the linear combination of atomic orbitals (LCAO) to the tightly bound transition term and clarify the relationship between the Tan–Bo model and SK/Harrison scaling law.

**(2)** The transition integrals, energy bands, and state densities of the isotropic Tan–Bo model, geometrically anisotropic model, and SK model were compared under equilibrium structures.

**(3)** We analyzed the band structures of both Hermitian symmetry-protected and fully non-Hermitian configurations systematically to elucidate the physical origin of the non-Hermitian imaginary component and its distribution along high-symmetry paths while simultaneously mapping the complex eigenvalues to their trajectories in the complex plane.

**(4)** The *f(k)* function is embedded within the Tan–Bo layer into the McCann Bernal bilayer model to verify the robustness of the $\Delta_K \approx 2\delta$ linear relationship for intralayer parameters.

**(5)** We analyzed the hybrid hopping scheme under different strain conditions and obtained the optimal *B* parameters for zigzag and armchair directions.

(6) The Hermitian Kubo formula is adopted to recover the universal Dirac scale and a TB van Hove peak blueshifted relative to the excitonic resonance. Thus, geometric anisotropy and hybrid strain generate dichroism qualitatively consistent with strained transmittance trends.

## 2. Principles and Calculation Methods

### 2.1 TB Model and Tan–Bo Geometric Transition

Under the atomic orbital basis, the π-band Hamiltonian is expressed as $\hat{H} = \sum_{i,j,\sigma} t_{ij} c^{\dagger}_{i\sigma} c_{j\sigma} + h.c.$, where $\langle i, j \rangle$ denotes the pair of nearest neighboring lattice points, $\sigma$ represents spin, and *h.c.* denotes the Hermitian conjugate term. Swapping the indices $i \leftrightarrow j$ and taking the complex conjugate $t^*$ do not involve repeated summation operations. The LCAO estimate yields $t_{ij} \approx V_{pp\pi} S_{ij}$ with an in-plane bond overlap integral satisfying $S_{ij} \sim \left(\frac{r_0}{r}\right)^2$. The tightly bound Hamiltonian for the two sublattices of graphene is given as

$$H(k) = \begin{pmatrix} 0 & f(k) \\ f^*(k) & 0 \end{pmatrix}, \quad f(k) = \sum_{j=1}^{3} t_j e^{-ik\cdot\boldsymbol{\delta}_j} \text{。}$$

This eigenvalue corresponds to $E_{\pm}(\mathbf{k}) = \pm |f(\mathbf{k})|$. At point K, *f(k)* = 0 yields the Γ Dirac point, i.e., $E = \pm 3|t|$, whereas at point M, $|E| \approx |t|$. The lattice constant *a* and the equilibrium bond length $r_0$.

The Tan–Bo model decomposes the transition integral ***t(dr)*** into a direction-dependent component $\phi(w)$ and an exponential decay factor that depends on the bond length as given below[23, 24].

$$t(dr) = \frac{Aw+B}{Cw+D} \cdot exp(-\frac{r-r_0}{\lambda}) = \phi(w) \cdot e^{-\frac{(r-r_0)}{\lambda}} .$$

The physical meaning of each symbol is as follows. ***dr*** is the bond vector (in Cartesian coordinates) pointing from the source atom to the target atom; the bond length $r = |dr| = \sqrt{X^2+Y^2+Z^2}$; and $w$ *is* the complex representation of the bond vector. For in-plane graphene (*Z* = 0), $w = X + iY$ is entirely determined by the bond orientation. $\phi(w) = \frac{Aw+B}{Cw+D}$ is the Möbius transformation, which maps the bond direction *w* to the transition amplitude and determines the variation of the transition integral with direction. The exponential decay factor $e^{-(r-r_0)/\lambda}$ describes the exponential decay of the transition integral with increasing bond length, where $r_0$. is the equilibrium bond length and $\lambda$ is the decay constant. The physical meanings of the Möbius parameters *A*, *B*, *C*, and *D* are presented in **Table 1**.

**Table 1 Physical meanings of Möbius parameters *A*, *B*, *C*, and *D***

| ***Parameter*** | ***Mathematical Position*** | ***Intuitive Effect*** | ***Effect*** |
|---|---|---|---|
| *A* | Molecule: *Aw* | The "fundamental transition" component in phase with *w* | The overall component varies synchronously with the chemical-bond orientation |
| *B* | Molecule: Constant Term | Apply *B*/*w*-type direction correction (without multiplying by *w*) | Isotropic when *B* = 0; The larger the *B* value, the greater is the difference in bond transitions. |
| *C* | Denominator：*Cw* | Direction Modulation Depth + Pole Position | Control the sensitivity of the denominator to direction; decide the poles simultaneously with *D* $w_p = -D/C$ |
| *D* | Denominator: Constant term | Denominator reference, pole shift | Set the pole position in conjunction with *C*; the form is simplest when *D* = 0 |

Pole: $Cw + D = 0 \Rightarrow$, $w_p = -D/C$. When $C \neq 0$, $\varphi(w)$ diverges at $w_p$; physically, $w_p$ must not to lie along any of the three nearest-neighbor bond directions $w_j$. When $C$ and $D$ are fixed, reducing $|C|$ increases the anisotropy of $|\phi(w)|$ with respect to the bond direction. Additionally, increasing $|C|$ renders $\phi(w)$ "flatter." $D$ shifts the pole from $w = 0$ ($D = 0$) to $w_p = -D/C$ and thus breaks the denominator symmetry of the triple bond and potentially opening the K-point gap. For the reverse bond property ( $B \neq 0$), $\phi(-w) = (-Aw + B)/(-Cw + D) = [\varphi(w)]^*$. Hence, $t(-d\mathbf{r}) \neq td\mathbf{r}^*$ in general, which constitutes the mathematical basis of non-Hermitian assembly.

In this study, we set $C = 1$ and $D = 0$ to simplify the Möbius transformation into $\phi(w) = A + \frac{B}{w}$ . The parameter $A$ determines the isotropic background (under equilibrium conditions, $A \approx -2.8\,\text{eV}$ ); the term $B/w$ introduces direction-dependent corrections; $B = 0$ corresponds to isotropy (standard tight binding); and $B \neq 0$ induces bond-dependent complex transitions and typically yields $t(-d\mathbf{r}) \neq t(d\mathbf{r})^*$ , which is the non-Hermitian behavior at the microscopic level.

This study employs fixed parameters, i.e., $A = -2.8$ eV, $C = 1$, $D = 0$, $r_0 = 1.42\,\mathring{\text{A}}$ . and $\lambda = 0.5\,\mathring{\text{A}}$. For example, the geometric preset sets $B = -3$ (where the negative sign only affects the imaginary-component symbol without affecting $|t|$ broadening), thus demonstrating direction dependence and non-hermiticity effects. Under the isotropic preset ($B = 0$), $\phi(w) = A$ , all three nearest-neighbor bond transitions were uniformly set to –2.8 eV($r = r_0$), which is consistent with both the standard TB and SK models.

The geometric preset is established with ($B$ = -3): $\phi(w) = -2.8 - \frac{3}{w}$ . Under the equilibrium bond length $r = r_0$ , the bond vectors of graphene's three nearest-neighbor bonds (in units of the lattice constant $a = 2.46\,\mathring{\text{A}}$) are expressed as follows.

$$w_1 = \frac{a}{2} + \frac{a}{2\sqrt{3}} i \approx 1.230 + 0.710i,\ w_2 = -\frac{a}{2} + \frac{a}{2\sqrt{3}} i \approx -1.230 + 0.710i,\ w_3 = -\frac{a}{\sqrt{3}} i \approx -1.420i$$

The corresponding transition integrals are as follows.

$t_1$ = -4.627+1.056$i$ (eV)，$t_2$ = -0.970+1.056$i$ (eV)，$t_3$ = -2.798-2.111$i$ (eV)。

At this $|t|$ point, the bandwidth is 3.31 eV. Given that $w$ changes to $-w$, $\phi(-w) = A - \frac{B}{w}$, and $\phi(w)^* = A + \frac{B}{w}$ under the reverse bond (where $A$ and $B$ are real numbers), the two values generally do not equal each other. Thus, $t(-d\mathbf{r}) \neq t(d\mathbf{r})^*$.

$C$ = 1 and $D$ = 0 are set based on physical principles. Although $C$ and $D$ can mathematically modify the strain-dependent curvature $\boldsymbol{t}(\mathrm{d}\boldsymbol{r})$ and energy bands in pristine graphene, they are subject to constraints. A value of $D \neq 0$ would open an electronvolt-scale bandgap without strain, which would contradict the experimental observation of closed K-point Dirac cones. $C$ serves primarily as a global scaling function that partially overlaps with the effects of simultaneously adjusting $A$ and $B$ (indicating parameter redundancy). When $B$ = 0, the model strictly matches the experimental data and thereby yields $B_{\mathrm{opt}}$ = 0 and minimizes the MSE. Thus, $C$ = 1 and $D$ = 0 represent both physical constraints and optimal parameter specifications, with physical modulation concentrated on $B$ (bond anisotropy) and strain hybrid effects. Parameter scans for $C$ and $D$ were conducted to clarify their functional roles and K-point closure conditions. In pristine graphene, $C = 1$ and $D = 0$ satisfy these physical constraints: the pole $w_p = -D/C$ lies at the origin, all three nearest neighbors $w_j$ exhibit identical $|Cw_j + D|$ values due to 120° rotational symmetry, thus maintaining closed K point configurations, and the $B$ term serves as a $1/w$-type directional correction near the origin that independently regulates triple-bond anisotropy. In low-symmetry environments such as molecular bonds, Moiré stacking AA regions, surface adsorbates, and defects/interfaces, the local coordination no longer obeys triple rotational symmetry, which allows $C$ and $D$ to vary freely. Thus, the pole position $w_p = -D/C$ describes resonance enhancement (with $|\boldsymbol{t}|$ significantly amplified on bonds near the pole) or mirror symmetry breaking along specific bond angles. At this stage, the correction near origin $B$ and the positioning of poles $C$/$D$ differ physically in their modulation mechanisms relative to the denominator. In these complex environments, the strain component preserves the hybrid transition definition described in Section 2.4, with $\boldsymbol{t}_{\mathrm{SK}}(\boldsymbol{r})$ maintained as a distance-dependent backbone consistent with experimental scales, whereas $C$/$D$/$B$ only introduces additional angular or local modulations. A more detailed discussion of parameters B, C, and D of the Tan-Bo model is provided below in that section.

### 2.2 Assembly of Unitary and Nonunitary Hamiltonians and Properties of Möbius Transformation

The forward transition structural factor $f_{\text{fwd}}(\boldsymbol{k})$ and backward transition structural factor $f_{\text{bwd}}(\boldsymbol{k})$ are defined as follows.

$$f_{\text{fwd}}(k) = \sum_{j=1} t(\boldsymbol{\delta}_j) e^{-ik\cdot\boldsymbol{\delta}_j}, \quad f_{\text{bwd}}(k) = \sum_{j=1} t(-\boldsymbol{\delta}_j) e^{ik\cdot\boldsymbol{\delta}_j} 。$$

The Hermitian mode enforcement $H_{BA} = f^*_{\text{fwd}}(k)$, i.e.,

$$H_{\text{H}}(k) = \begin{pmatrix} 0 & f_{\text{fwd}}(k) \\ f^*_{\text{fwd}}(k) & 0 \end{pmatrix}.$$

According to the standard TB convention, the Dirac cone at point K remains closed in equilibrium.

The non-Hermitian mode directly employs the originally defined $f_{\text{fwd}}(\boldsymbol{k})$ and $f_{\text{bwd}}(\boldsymbol{k})$.

$$H_{\text{NH}}(k) = \begin{pmatrix} 0 & f_{\text{fwd}}(k) \\ f_{\text{bwd}}(k) & 0 \end{pmatrix}.$$

In $t(-\boldsymbol{\delta}_j) = t(\boldsymbol{\delta}_j)^*$ $H_{\text{NH}} = H_{\text{H}}$ the geometric Tan–Bo model where $B \neq 0$ and $H_{\text{NH}} \neq H^\dagger_{\text{H}}$, the eigenvalues typically exhibit nonzero imaginary components. This occurs because $\varphi(-w) = A - B/w$ does not equal $[A + B/w]^*$ (with $C = 1$ and $D = 0$). The value Im($E$) corresponds to the noninterchangeable TB model instead of the quasi-particle lifetime induced by dissipation.

By projecting the bond vector onto the complex plane $w = X + iY$, the transformation $\phi(w) = \dfrac{Aw+B}{Cw+D}$ exhibits the following characteristics.

When $B \neq 0$, dipole anisotropy of the $1/w$ type is induced, which results in distinct transition integrals for the three nearest-neighbor bonds. The denominator $Cw + D$ introduces a pole (to prevent alignment with physical bond directions), thereby enhancing directional sensitivity nonlinearity. The parameters $A$, $B$, $C$, and $D$ independently control various geometric characteristics to facilitate adjustment of anisotropy intensity.

Natural origin of nonreciprocity: Under the simplified assumptions of this study, $\phi(w)=\frac{A+B}{w}$. For the reverse bonds, we used $w\rightarrow -w$, $\phi(-w)=A-\frac{B}{w}$, and $\phi(w)^*=A+\frac{B}{w}$ ($A$ and $B$ are real numbers). This equality holds only when $B$ = 0. Thus, a nonzero value of $B$ yield $t(-d\mathbf{r})\neq t(d\mathbf{r})^*$, which results in complex energy bands in non-Hermitian assemblies. Consequently, the Möbius transformation serves as a concise and effective tool to investigate nonreciprocal TB models.

**2.3 Parameter Calibration and B-parameter Calibration**

As a reference for distance scaling, we adopted the Harrison form (*ppπ* channel)

$$t_{\mathrm{SK}}(r)=V_{pp\pi}^{0}\left(\frac{r_0}{r}\right)^2,$$

where $V_{pp\pi}^{0}=-2.8\ \mathrm{eV}$ and $r_0=1.42\,\mathrm{\mathring{A}}$. With the Tan–Bo model parameters fixed at $A$ = −2.8, $C$ = 1, and $D$ = 0, the objective function is minimized within the narrow interval $B$ ∈ [−0.05,0.05].

$$L(B)=\mathrm{MSE}(t_j^{TanBo},t_j^{SK})+5000\,\Delta_K(B)^2,$$

In the equation above, $\mathrm{MSE}=\frac{1}{3}\sum_{j=1}^{3}\left|t_j^{TanBo}-t_j^{SK}\right|^2$ and $\Delta_K$ (B) represents the K-point bandgap under the Hermite mode. In an equilibrium structure, $B_{\mathrm{opt}}=0$, thus yielding $\mathrm{MSE}=2.18\times10^{-7}\ \mathrm{eV}^2$ and $\Delta_K\approx2.7\times10^{-15}\ \mathrm{eV}$. At this point, the Tan–Bo isotropy ($B$ = 0) matches precisely with the standard TB/SK model.

The abovementioned parameter $L(B)$ is used solely for SK calibration under equilibrium geometry with its search interval $b_{\mathrm{bounds}}$ = (−0.05,0.05), deliberately excluding $B$ = −3. The geometric preset $B$ = −3 represents an independent manual parameterization employed to demonstrate strong bond anisotropy ($\Delta|t|\approx 3.31$ eV) and the non-Hermitian effect; it is not claimed to be the global optimum for $L$(B). Under a 5% zigzag strain, if the full Tan–Bo model (non-hybrid) is employed with $B_{opt}\approx 0$, then the MSE values for the Tan–Bo $t_j$ and $t_{SK}$ are approximately $4.5\times10^{-3}\ \mathrm{eV}^2$. This implies that a single scalar parameter B cannot simultaneously fit all three differentiated strain bonds. Therefore, the hybrid scheme described in Section 2.4 must be adopted: $\boldsymbol{t_j}=(\varphi/\varphi_{eq})\times \boldsymbol{t_{SK}}(\boldsymbol{r_j})$, where the Harrison/Pereira relation accounts for distance dependence and the Möbius ratio governs only the angular modulation.

### 2.4 Hybrid Transition Under Strain

In strain studies, substituting the complete Tan–Bo index formula $t = \varphi(w) \exp[-(r - r_0)/\lambda]$ into the K-point gap calculation yields a nonphysical gap on the order of electronvolts under strain. The only physically meaningful definition comparable to experimental or DFT results is that of hybrid transitions.

$$t_j^{hybrid}(\varepsilon) = \begin{cases} t_{SK}(r_j), & B = 0 \\ \left[\dfrac{\phi(w_j)}{\phi_{eq}(w_j)}\right] \times t_{SK}(r_j), & B \neq 0 \end{cases}.$$

When $B = 0$, $t_j^{hybrid} = t_{SK}(r_j)$ (the Pereira exponential hopping can be used as the basis), which strictly follows the Harrison ( $t_j^{hybrid} = t_{Harrison}(r_j) = V_{\mathrm{pp}\pi} \cdot (r_0/r_j)^2$ )/Pereira ( $t_j^{hybrid} = t_{Pereira}(r_j) = t_0\ exp[-\beta(l_j - 1)]$ ) distance law. The Möbius $\dfrac{\phi(w_j)}{\phi_{eq}(w_j)}$ ratio has no effect, and the system exhibits no additional angular modulation.

When $B \neq 0$, $t_j^{hybrid}(\varepsilon) = \left[\dfrac{\phi(w_j)}{\phi_{eq}(w_j)}\right] \times t_{SK}(r_j)$, and $\phi_{eq}(w_j)$ is the Möbius factor for balancing bond $j$. The Harrison scaling determines the bond-length dependence upon which the Möbius geometric factor $\phi(w)/\phi_{\mathrm{eq}}$ superimposes a controllable anisotropy of the three-bond |t|. At equilibrium ($r = r_0$, no strain), $\phi/\phi_{\mathrm{eq}}$ and the exponential factor are 1, thereby reverting to the original Tan–Bo hopping (consistent with Section 2.3). Two definitions of the K-point gap ( $k_{\mathrm{gap}}/k_{\mathrm{gap(re)}}$) are given as follows. Hermitian: $\Delta_K = |E_+(K)\text{-}E_-(K)|$; non-Hermitian: $\Delta_K(\mathrm{Re}) = |\mathrm{Re}E_+(K)\text{-}\mathrm{Re}E_-(K)|$. The complex modulus difference $|E_+\text{-}E_-|$ is explicitly excluded in the calculation because it has no experimental counterpart in strain scanning and would introduce nonreciprocal artifacts.

### 2.5 Bernal Bilayer and Numerical Methods

McCann's four-band model was adopted with the basis set ($A_1$, $B_1$, $A_2$, $B_2$). The Hamiltonian is expressed as

$$H_{\mathrm{bilayer}} = \begin{vmatrix} \delta & f(k) & 0 & \gamma_1 \\ f^*(k) & \delta & 0 & 0 \\ 0 & 0 & -\delta & f(k) \\ \gamma_1 & 0 & f^*(k) & -\delta \end{vmatrix},$$

where $H_{\mathrm{mono}}(k) = \begin{pmatrix} 0 & f(k) \\ f^*(k) & 0 \end{pmatrix}$ (monolayer, using Tan–Bo), $\gamma_0 \approx -2.8$ eV (contained in $f$), and $\gamma_1 = -0.39$ eV. Here, $\gamma_1$ denotes the vertical coupling between $B_1$ and $A_2$ vertical sites, while interlayer symmetry breaking is introduced via sublattices $+\delta$ and $-\delta$ on layers 1 and 2 (eV), respectively. Under zero bias, $\delta = 0$; when $\delta = 50$ meV, $\Delta_K = 0.1\,\mathrm{eV}$, thus satisfying $\Delta_K \approx 2\delta$. The band structure calculation path is Γ → K → M → Γ, with 360 K-points per segment and tick positions consistent with those reported in standard graphene literature. The density of states (DOS) is computed using a uniform 2D $k$-mesh, with Gaussian broadening $\sigma = 0.06$–$0.08$ eV. All bilayer calculations shared the same set of Tan–Bo intralayer parameters as the monolayer calculations to ensure internal consistency in the iso/geo comparisons.

### 2.6 Strain Application and Calculation of the Hopping Integral

We define the strain tensor $\boldsymbol{F} = I + \varepsilon_{\mathrm{tensor}}$. For the two typical edge configurations, uniaxial tensile strain was applied as follows.

Zigzag edge: stress was applied along the x-direction with Poisson's ratio $\nu \approx 0.165$, and the strain tensor expressed as

$$\varepsilon_{\mathrm{tensor}} = \begin{bmatrix} \varepsilon & 0 \\ 0 & -\nu\varepsilon \end{bmatrix}.$$

Armchair edge: stress applied along the y-direction and the strain tensor expressed as

$$\varepsilon_{\mathrm{tensor}} = \begin{bmatrix} -\nu\varepsilon & 0 \\ 0 & \varepsilon \end{bmatrix}.$$

The rotated components are ($\varepsilon_{11}$, $\varepsilon_{22}$, $\varepsilon_{12}$), whereas the deformed bond vectors are $\boldsymbol{dr}_{j(\varepsilon)}$: $\boldsymbol{dr}_{j(\varepsilon)} = F \cdot dr_{j^0}$ (acting only on the $xy$ components, $Z$ remains 0) and $\boldsymbol{r}_j = |\boldsymbol{dr}_{j(\varepsilon)}|$. Generally, these three bond lengths are no longer equal. For example, under 5% zigzag tension, two bonds are elongated and one is slightly shortened; the specific values are determined by $\boldsymbol{F}$ and $dr_j^0$. To calculate the hopping integral subsequently, a dimensionless bond length must be introduced. Following the work of Pereira et al., the dimensionless quantities corresponding to the three bonds are defined as $l_j$（$j = 1,2,3$）, as shown below.

$$l_1 = 1 + 0.75\varepsilon_{11} - 0.5\sqrt{3}\varepsilon_{12} + 0.25\varepsilon_{22},$$
$$l_2 = 1 + \varepsilon_{22},$$
$$l_3 = 1 + 0.75\varepsilon_{11} + 0.5\sqrt{3}\varepsilon_{12} + 0.25\varepsilon_{22},$$

Where $l$ denotes the components of the strain tensor after rotation. In this work, we adopt two schemes to calculate the nearest-neighbor hopping integrals.

A. SK:

$$\boldsymbol{t}_{\mathrm{SK}}(\boldsymbol{r}) = V_{pp\pi} \cdot (r_0/r)^2,\ V_{\mathrm{pp\pi}} = -2.8\ \mathrm{eV},\ r_0 = 1.42\ \text{Å}.$$

B. Pereira exponential hopping:

$$t_j^{\mathrm{Pereira}} = t_0 \cdot \exp[-\beta \cdot (l_j - 1)],$$

where $t_0 = $ -2.8 eV and $\beta = 3.37$. The hopping integrals $t_1, t_2, and\ t_3$ of the three bonds are mapped to the three bond vectors based on the corresponding bond indices. When the geometric parameter $B = 0$, the hopping integral is expressed as $t_j = t_{\mathrm{SK},j}$ (or $t_{\mathrm{Pereira},j}$) without additional Möbius modulation.

The reference Möbius factors for the unstrained equilibrium configuration must be precalculated only when $B \neq 0$. For the equilibrium bond vectors $dr_j^0$, we define $w_{j^o} = (X_{j^o} + iY_{j^o}) / (1 - Z_{j^o})$ and then calculate $\varphi_{eq}(w_{j^o}) = (A \cdot w_{j^o} + B) / (C \cdot w_{j^o} + D)$, denoted as $ref_j$. In the equilibrium state, the three bonds correspond approximately to

$$w_1 \approx 1.230 + 0.710i,$$

$$w_2 \approx -1.230 + 0.710i,$$

$$w_3 \approx -1.420i.$$

These reference values are calculated once during model initialization and remain unchanged throughout the strain simulations. For the strained bond vectors $dr_{j(\varepsilon)}$, we similarly compute

$$w_j = (X_j + iY_j) / (1 - Z_j),$$

$$mob_j = \varphi(w_j) = (A \cdot w_j + B) / (C \cdot w_j + D),$$

The hopping integral ultimately used for non-Hermitian assembly is given as

$$t_{j(\varepsilon)} = (mob_j / ref_j) \times t_j^{\mathrm{Pereira}}.$$

A summary of the treatment under different conditions is presented in **Table 2**.

**Table 2** Summary of hopping parameters and Möbius modulation under varying strain and magnetic field conditions.

| Condition | Result |
|---|---|
| No strain, $r_j = r_0$ | $mob_j = ref_j \rightarrow t_j$ = Tan–Bo hopping |
| Strained, $B = 0$ | $mob = A \cdot w$, $ref = A \cdot w^0$; however, either $t_{SK}$ or $t_{Pereira}$ is used directly, without ratio modulation |
| Strained, $B \neq 0$ | Distance and angular effects are reflected by the Pereira-type hopping and the relative Möbius modulation, respectively |

Reverse bonds (for non-Hermitian assembly): for $-\boldsymbol{dr}_j$, bond hopping($-\boldsymbol{dr}_j$, $bond_{idx} = j$) is invoked in the same manner, generally $t(-\boldsymbol{dr}) \neq t(\boldsymbol{dr})^*$.

In the non-Hermitian Hamiltonian assembly, the forward and backward hopping phase factors are defined as follows.

$$f_{\text{fwd}}(k) = \sum_{j=1} t_j(\varepsilon) e^{-ik \cdot dr_j}, \quad f_{\text{bwd}}(k) = \sum_{j=1} t_j(-dr_j) e^{ik \cdot dr_j}.$$

In the Hermitian case,

$$H(\boldsymbol{k}) = \begin{bmatrix} 0 & f_{\text{fwd}} \\ f_{\text{fwd}}^* & 0 \end{bmatrix},$$

and the eigenvalues are expressed as $E = \pm |f_{\text{fwd}}|$.

In the non-Hermitian case (for displaying the imaginary component $\text{Im}(E)$),

$$H(\boldsymbol{k}) = \begin{bmatrix} 0 & f_{\text{fwd}} \\ f_{\text{bwd}} & 0 \end{bmatrix}.$$

In this case, the eigenvalues are generally complex, with their real and imaginary components respectively denoted by $\text{Re}(E)$and $\text{Im}(E)$. This section describes the complete workflow for strain-dependent non-Hermitian band-structure calculations. In this study, the strain magnitudes were set as 3%, 6%, 12%, and 24%, with geometric parameters of 0, −0.75, −1.5, and −3. Calculations were performed for both the zigzag and armchair edge configurations, and variations in the real and imaginary components of the energy eigenvalues were compared.

### 2.7 Optical conductivity

The real interband conductivity is obtained from the Kubo–Greenwood formula using velocity operators $v_\alpha = \partial H/\partial k_\alpha$ on the Hermitian Tan–Bo Hamiltonian, and is normalized to the graphene universal conductivity

$$\sigma_0 = \frac{e^2}{4\hbar} = \frac{\pi e^2}{2h}, \ h = 2\pi\hbar,$$

i.e. the mid-infrared Dirac sheet conductivity $\pi e^2/2h$.[25] Non-Hermitian assemblies are excluded from optical calculations: Im E is a structural nonreciprocal-assembly diagnostic and is not treated as a quasiparticle lifetime or $\hbar/2\tau$.

## 3. Results and Discussion

### 3.1 Balancing Structural Transition and SK Calibration

To begin, we benchmark the parameter behavior of the Tan–Bo model under the equilibrium geometry. **Table 3** summarizes the transition integrals of the three nearest-neighbor bonds under an equilibrium structure ($r_0 = 1.42$ Å). The isotropic Tan–Bo ($B = 0$), SK (Harrison), and LCAO models estimate $|t| \approx 2.8$ eV, which is consistent with the standard TB method. Fitting to the SK model yields $B_{\mathrm{opt}} = 0$, $MSE = 2.18\times10^{-7}$, and K-point bandgap $< 2.66\times10^{-15}$ eV. This result demonstrates that the SK model is equivalent to the isotropic Tan–Bo model for equilibrium graphene.

We now turn to the geometric preset (B=−3), which introduces substantial bond-directional modulation. The magnitudes of the three bonds were 4.746, 1.434, and 3.505 eV, whereas the bonds were $t_1 = -4.627 + 1.056i$, $t_2 = -0.970 + 1.056i$, and $t_3 = -2.798 - 2.111i$ eV ($\Delta|t| = 3.31$). Despite the pronounced differences in bond direction, the K-point bandgap remained at $0$ eV after Hermitian symmetrization, and the Dirac node remained closed.

**Table 3** Geometric presets ($B = -3$): Detailed data for three balanced bonds

| Bond | $w_j$ | $t_j$(eV) | $\|t_j\|$ |
|---|---|---|---|
| 1 | $1.230+0.710i$ | $-4.627+1.056i$ | 4.746 |
| 2 | $-1.230+0.710i$ | $-0.970+1.056i$ | 1.434 |
| 3 | $-1.420i$ | $-2.798-2.111i$ | 3.505 |

**Figure 1** illustrates the abovementioned bond-dependent effects. The Tan–Bo hopping integrals of the three nearest-neighbor bonds were no longer equal under forward and reverse bond vectors. The hopping magnitudes of bonds 1 and 2 were swapped, while the sign of the imaginary component of bond 3 switched between the forward and reverse directions, in sharp contrast to the SK model with a single $|t| \approx 2.8$ eV. The data show that $\max|t(-dr)-t(dr)^*| = 3.66$ eV; this nonreciprocal deviation is the direct microscopic origin of $\mathrm{Im}(E) \neq 0$ in subsequent non-Hermitian assemblies. However, in the Hermitian regime, the six hexagonal symmetry constraint enforced $f(K)=0$; therefore, the opening of the Dirac point is not necessarily due to the anisotropy of $|t|$.

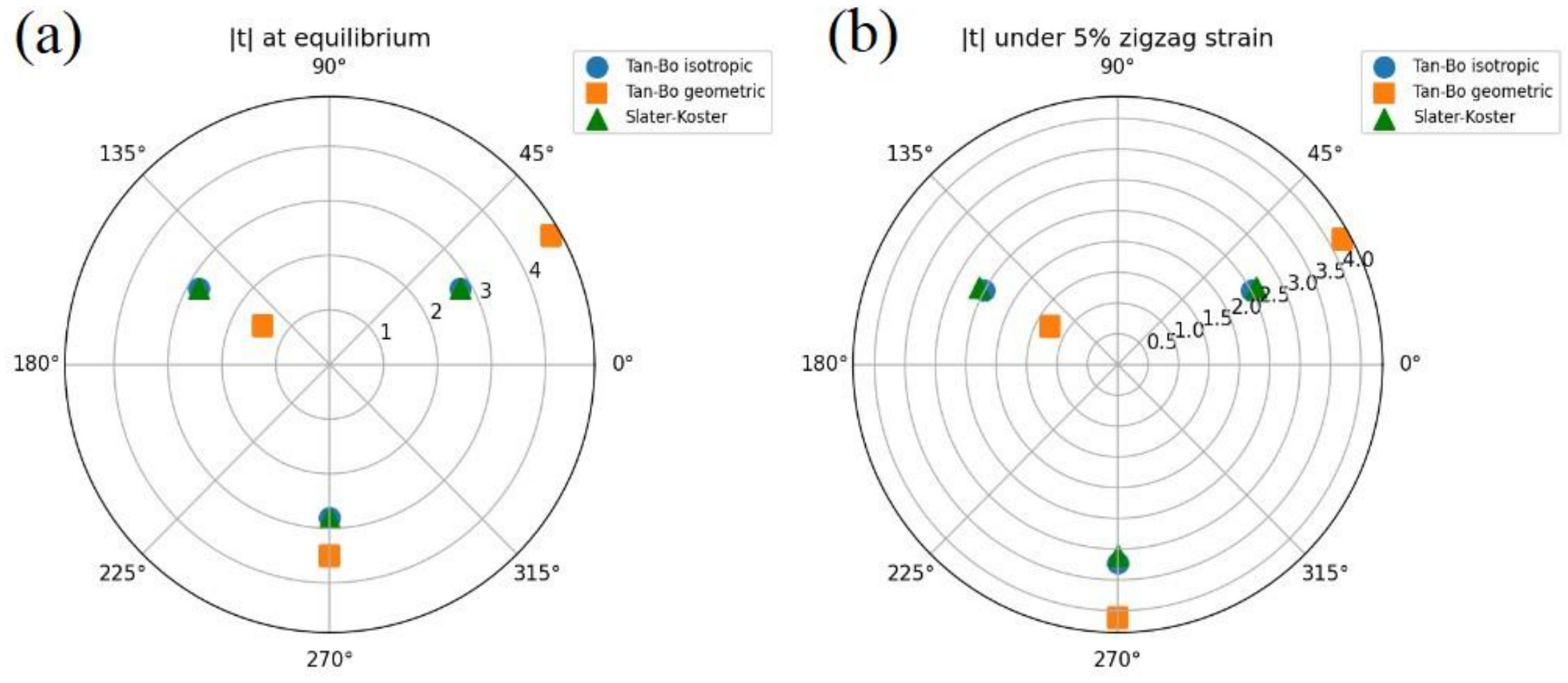

**Figure 1** Polar-coordinate comparison of $|t|$ for (a) equilibrium structure and (b) structure under 5% zigzag tensile strain.

### 3.2 Band structure and DOS

Under Hermitian symmetrization, the geometric anisotropy of the Tan–Bo model does not alter the Dirac physics at the K-point. However, it slightly modifies the dispersion relations at the M-point and in the high-energy region. The DOS maintains the V-shape characteristic of a semimetal near-zero energy, with the van Hove peaks shifting in response to the angular modulation of $|f(k)|$. **Figure 2a** shows a comparison of the band structures of the standard TB model and the geometric Tan–Bo (Hermitian, $B=-3$) model along the path $\Gamma \to \mathrm{K} \to \mathrm{M} \to \Gamma$. Both models formed Dirac cones at the K-point ($E \approx 0$), with the valence band top and conduction band bottom located at the $\Gamma$ point, i.e., $\pm 8.395$ eV. Similar characteristics were observed for the overall dispersion profiles. The data show that at the M-point, the energies were $\pm 2.80$ and $\pm 3.20$ eV for the TB $(\text{iso})$ and geometric (geo) models, respectively, with a difference of approximately 0.40 eV. The maximum deviation of $\mathrm{Re}(E)$ along the entire path was approximately 0.43 eV. This indicates that under strong bond-directional modulation, the Hermitian Dirac cone remains protected by the hexagonal symmetry, and the upward shift of the M point energy originates from the angular variation of $|f(k)|$ instead of gap opening at the K-point.

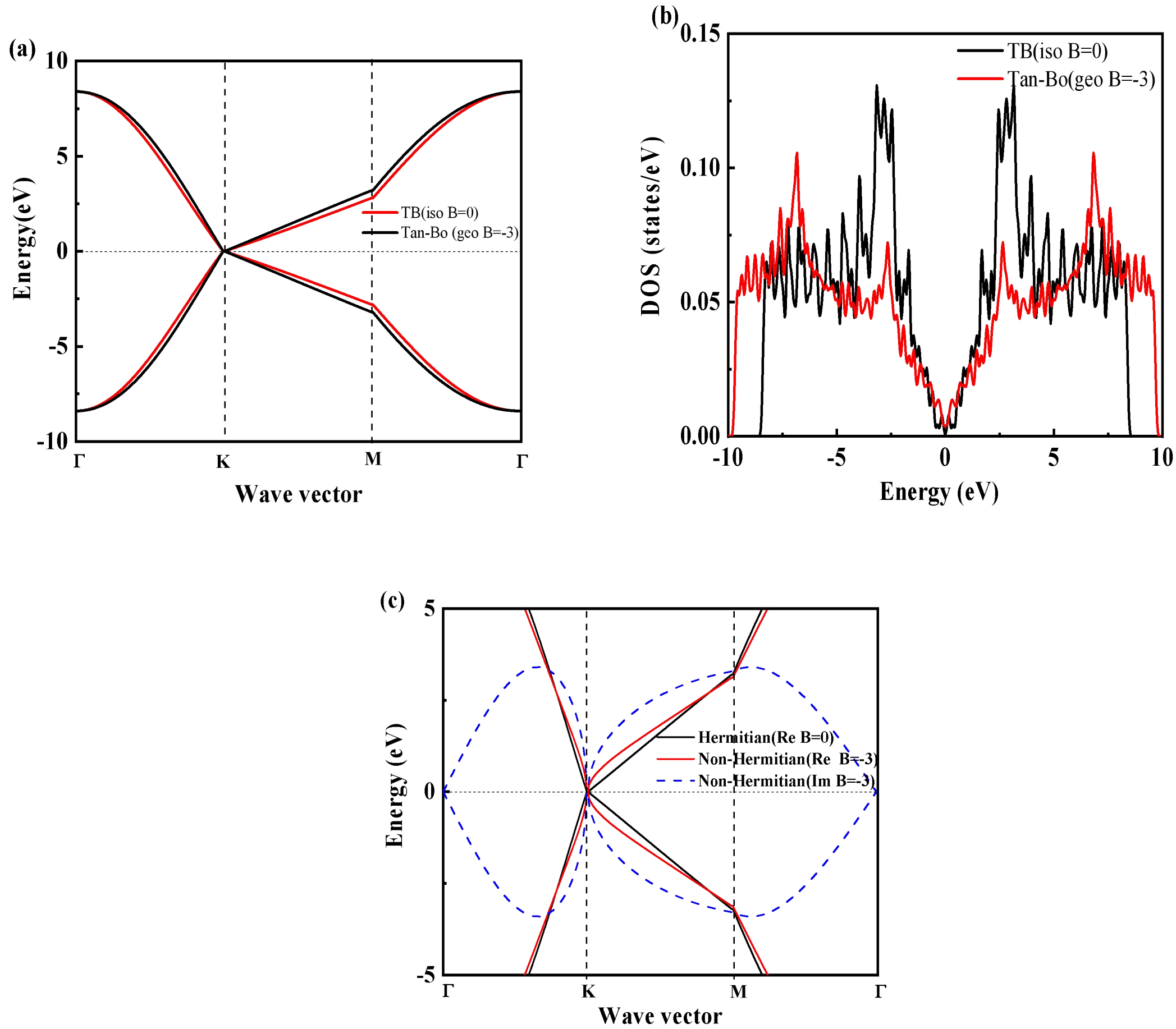

**Figure 2a** Comparison of band structure along Γ–K–M–Γ between standard TB (iso) and the geometric Tan–Bo ( $B$ = −3) models. (b) Density of states for Tan–Bo isotropic (TB) and geometric Tan–Bo models. (c) Tan–o model's Hermitian vs. non-Hermitian band structures

The density of states provides complementary insight. **Figure 2b** shows the DOS for the isotropic (TB) and geometric (Hermitian) models. Both curves were symmetric about $E = 0$ , and the DOS approached zero at the Dirac point. The isotropic model exhibited a main DOS peak of 0.131 states/eV at $|E| = 3.16$ eV, with a bandwidth of approximately $\pm 8.5$ eV. Meanwhile, the geometric model exhibited a broader bandwidth of up to $\pm 9.8$ eV (with a more extended high-energy tail) and a DOS of 0.106 states/eV at $|E| = 6.83$ eV. The DOS was higher in the intermediate-energy region for the isotropic model, whereas the high-energy tail was wider for the geometric model, which is consistent with the upward shift of the *M*-point energy shown in **Figure 2a**. This reflects the reshaping of the van Hove structure by the angular modulation of $|f(k)|$.

We next consider the fully non-Hermitian assembly. Under a full non-Hermitian assembly, $\mathrm{Im}(E)$ vanishes at high-symmetry points and reaches the electronvolt scale along the $K-M$ segment. The $\mathrm{Re}(E)$ dispersion remained similar to that of the Hermitian case, whereas the imaginary component characterizes the k-space distribution of nonreciprocal hopping and should not be equated with the quasiparticle lifetime or dissipation in open systems. **Figure 2c** shows the geometric Tan–Bo non-Hermitian band structure for $\mathrm{Re}(E)$ and $|\mathrm{Im}(E)|$: at $\mathrm{Re}(E)$, $K$-point, the bands crossed, whereas at $|\mathrm{Im}(E)|$, $\Gamma$, and *K*, they were strictly zero but increased significantly along $K-M-\Gamma$, with a peak value on the order of electronvolts. The gap at $\mathrm{Re}(E)$ is 0 eV, where $E=-3.115+3.284i$ and $3.115-3.284i$ eV, and the path is $\max|\mathrm{Im}(E)|=3.407$. The imaginary-component peak coincides with the region of maximal nonreciprocal deviation in bond 3, which should be interpreted as a nonreciprocal TB assembly effect. Re(E) at Γ and K is ±8.395 eV and 0, respectively, whereas at M, the non-Hermitian *R*e(*E*) is ≈ ±3.12 eV (Hermitian geo: ±3.20 eV). The k-space mismatch between Re and Im indicates that non-Hermitian modifications primarily alter the hopping phase structure instead of shifting the overall Fermi level, and that $\mathrm{Re}(E)$ can still serve as the primary band-dispersion quantity. *B* primarily controls the noncommutative strength instead of the Dirac-point position. $B=0$ serves as a sufficient condition to eliminate the non-Hermitic imaginary component.

### 3.3 Bernal bilayer graphene

The Tan–Bo intralayer *f*(***k***) was embedded into the McCann four-band model（$\gamma_1=-0.39$ eV）. The iso and geo calculations yielded consistent conclusions for the bilayer topology at the K-point and under zero perpendicular bias. The geo calculation primarily elevated the energy at the M-point and along the k-path. At zero bias ($\delta=0$), the two middle bands coincided at the K-point, with a bandgap $<10^{-29}$ eV. Under perpendicular bias, the opening gap satisfied $\Delta_K=2\delta$. The bilayer DOS at the ±2.8 eV van Hove peaks was lower than that of the monolayer, and the fourfold band overlap near zero energy rendered the spectrum more complex. Comparing the monolayer's two-band model with the Bernal bilayer's four band model (**Figure 3a**), the monolayer exhibited linear band crossings at the K-point, whereas the bilayer showed parabolic touching of the two middle bands at the K point, with splitting induced by the interlayer hopping parameter $\gamma_1$. The interlayer splitting was visible at the M-point, and the intralayer energy at the M-point in the geo calculation was slightly higher. The iso and geo calculations agreed precisely at the K-point. However, a slight discrepancy appeared along the Γ–K path near K, without altering the four-band topology of the bilayer. Notably, $|f(k)_{\text{mono}} - f(k)_{\text{blg}}| < 10^{-10}$ eV. Furthermore, replacing only the intralayer $f(\boldsymbol{k})$ in the Tan–Bo model ensured that the McCann interlayer physics remained dominant. **Figure 3b** illustrates the variations in the K-point bandgap with the layer bias for both the iso and geo intralayer models. Data from the two plots overlapped completely with $\Delta_K$ increasing strictly linearly with $\delta$. At $\delta = 50$ meV, $\Delta_K = 0.100$ eV, whereas at $\delta = 120$ meV, $\Delta_K = 0.240$ eV. The slope was $\mathrm{d}\Delta_K / \mathrm{d}\delta = 2.0 \times 10^{-3}$ eV/meV, $\Delta_K \approx 2\delta$. These results are consistent with the McCann model in the limit of zero $\gamma_4$ and is independent of the intralayer $B$ parameters.

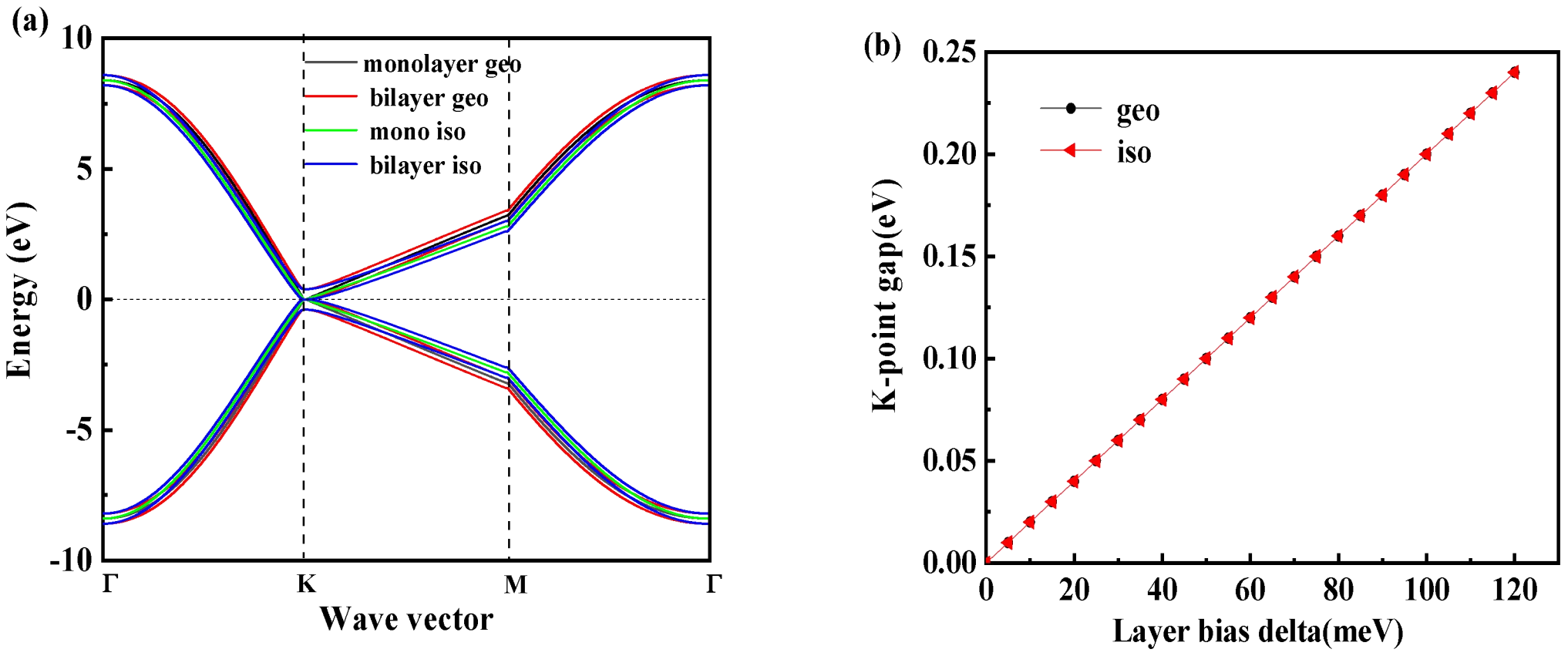

**Figure 3** Bernal bilayer band structure and K-point gap–bias relationship: (a) Monolayer 2-band vs. Bernal bilayer 4-band; (b) K-point gap ΔK as a function of interlayer bias δ—comparison between iso and geo intralayer models.

**Figure 4** shows a comparison of the single- and double-layer DOS at $E = 0$, where the single-layer peak is at −2.755 eV with 0.186 states/eV and the double-layer peak is at

$|E| = 2.6$

+2.988 eV with 0.137 states/eV. The double-layer spectrum exhibited denser oscillations in the region where $|E| < 5$ eV, reflecting four-band superposition. The isolated DOS of the bilayer shows a bimodal distribution at the electronvolt and 3.0 eV levels, with a DOS ≈ 0.003 at $E = 0$ arising from a Gaussian width σ of 0.06 eV instead of being strictly zero. A main peak emerged at |E| ≈ 7 eV, where both the monolayer and bilayer exhibited a DOS of approximately 0.088. The single-layer DOS was approximately 0.095 at ~6.8 eV. Although the bilayer DOS remained lower than that of the monolayer, its peak position differed significantly from that of the single layer (≈3 eV), and the geo parameter altered the overall energy scale of the spectrum.

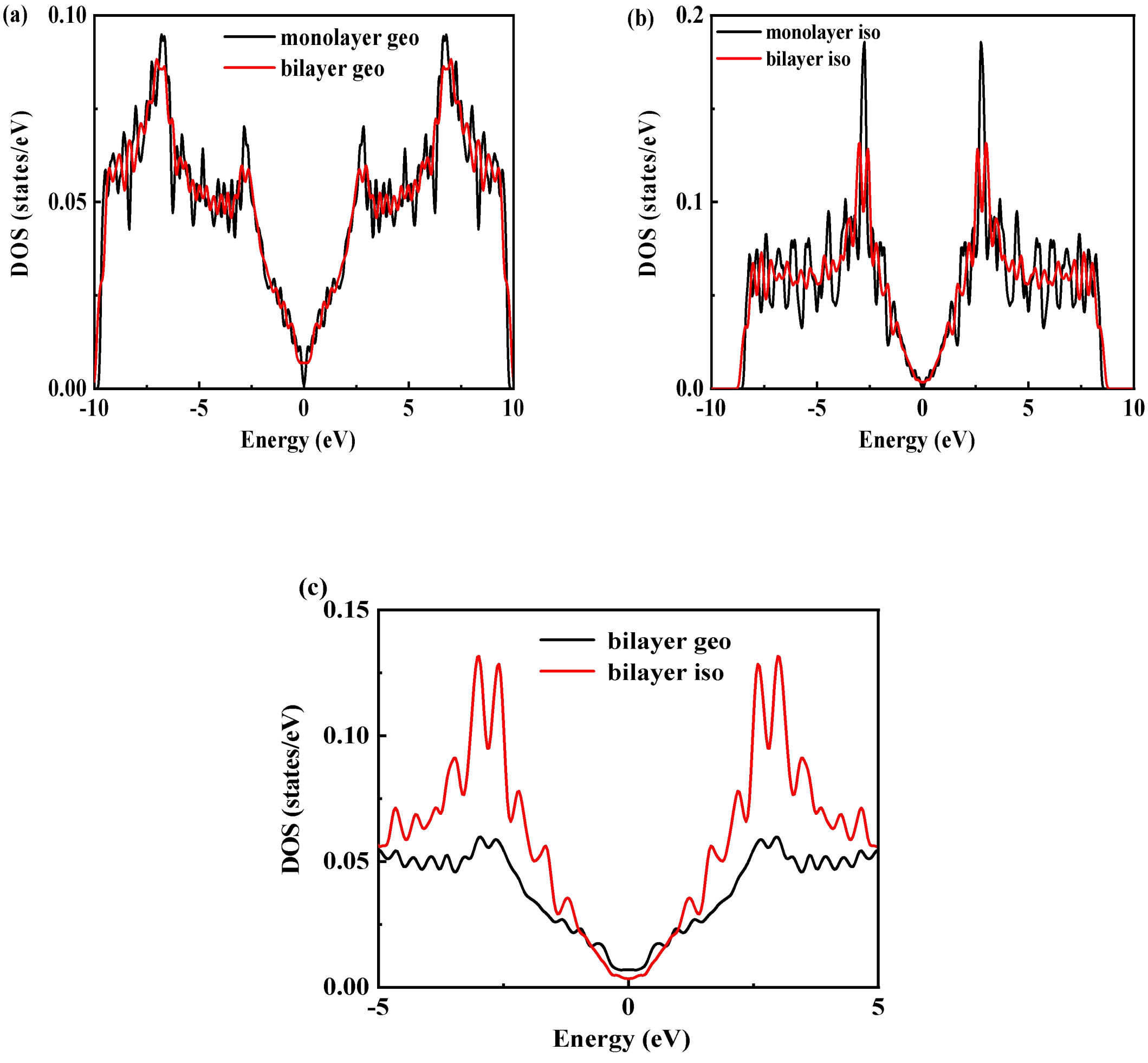

**Figure 4** (a) Tan–Bo geo monolayer vs. bilayer DOS; (b) Tan–Bo iso monolayer vs. bilayer DOS; (c) Tan–Bo iso monolayer ($B = 0$) vs. Tan–Bo geo ($B = -3$) bilayer DOS.

## 3.4 Band Structure Under Strain

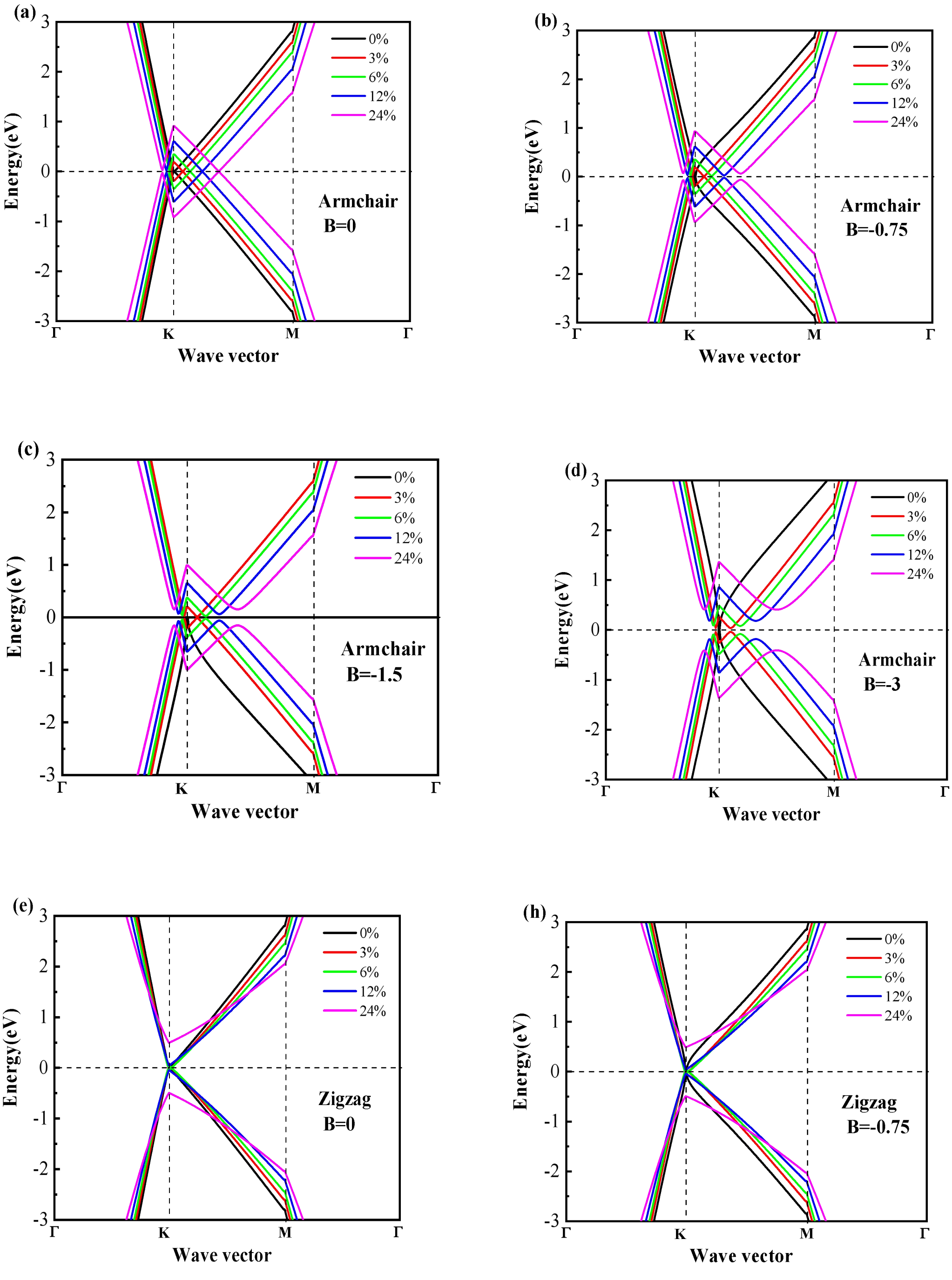

(a)
Energy(eV)
Wave vector
Γ K M Γ
0% 3% 6% 12% 24%
Armchair
B=0
(b)
Energy(eV)
Wave vector
Γ K M Γ
0% 3% 6% 12% 24%
Armchair
B=-0.75
(c)
Energy(eV)
Wave vector
Γ K M Γ
0% 3% 6% 12% 24%
Armchair
B=-1.5
(d)
Energy(eV)
Wave vector
Γ K M Γ
0% 3% 6% 12% 24%
Armchair
B=-3
(e)
Energy(eV)
Wave vector
Γ K M Γ
0% 3% 6% 12% 24%
Zigzag
B=0
(h)
Energy(eV)
Wave vector
Γ K M Γ
0% 3% 6% 12% 24%
Zigzag
B=-0.75

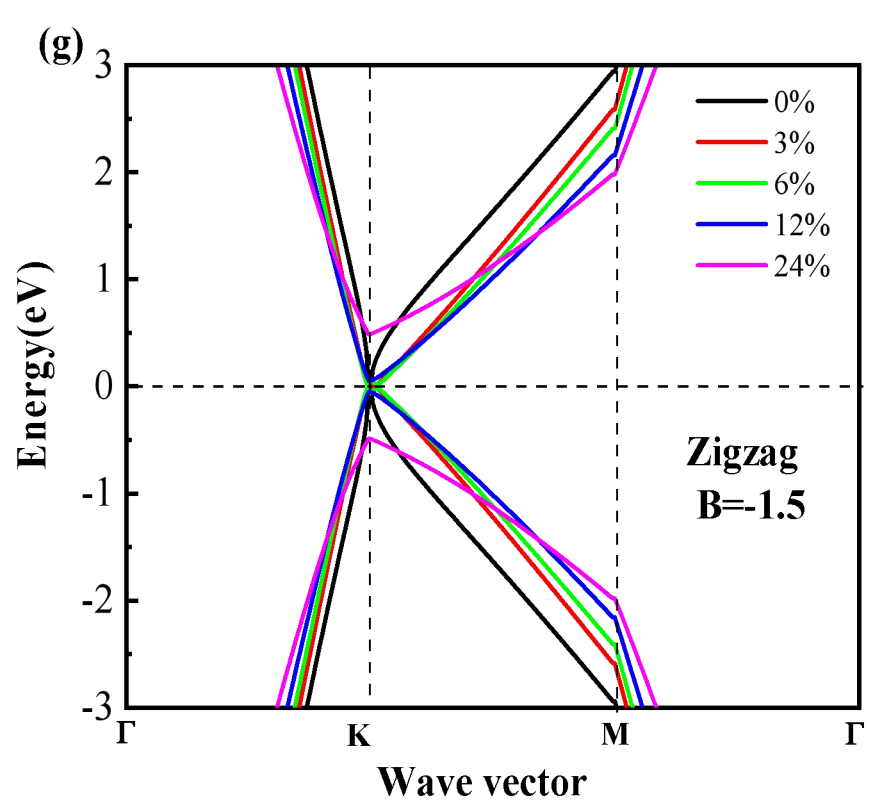

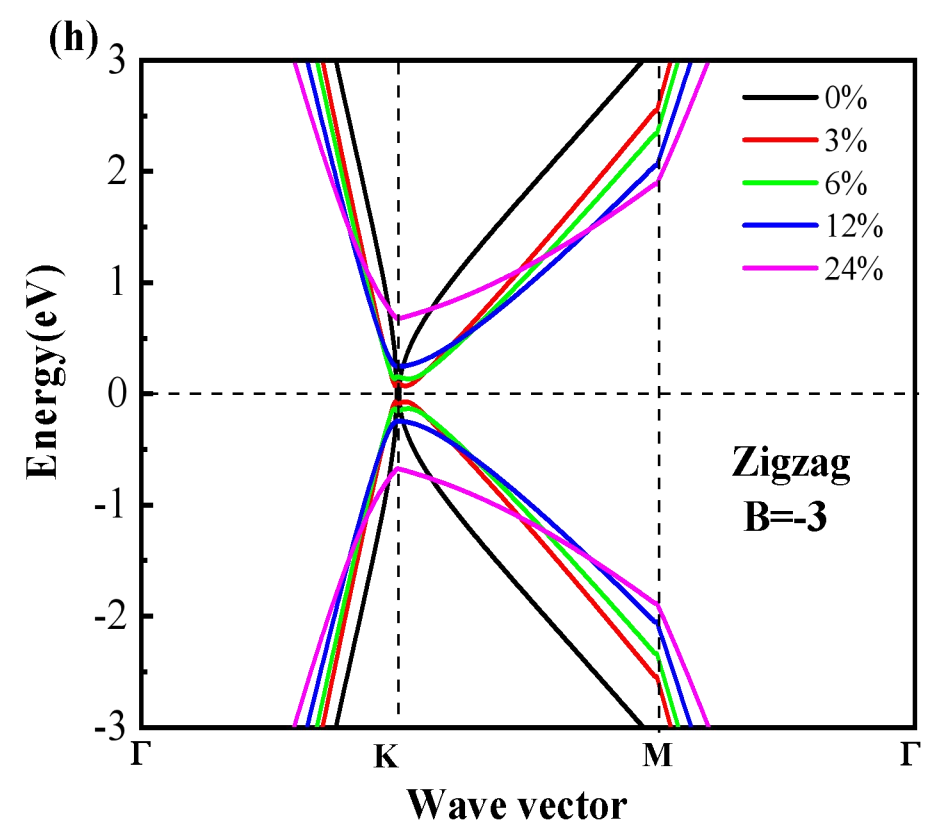

**Figure 5** Geo Tan–Bo non-Hermitian (NH) band structure for zigzag and armchair edge configurations under different strains $\varepsilon$ (3%, 6%, 12%, and 24%) and geometric parameters $B$ (0, −0.75, −1.5, and −3).

Finally, we investigate how uniaxial strain interacts with the Möbius angular correction. The non-Hermiticity of the geo Tan–Bo model arises from $\boldsymbol{t}(-d\boldsymbol{r}) \neq \boldsymbol{t}(d\boldsymbol{r})^*$, with $B$ controlling the Möbius-type angular correction. Using the geometric non-Hermitian Tan–Bo model, we systematically analyzed the band structure of graphene under strains ranging from 3% to 24% and $B$ ranging from 0 to −3, as shown in the **Figure 5**. The zigzag and armchair configurations exhibited pronounced anisotropy. Upon introducing $B \neq 0$, the bandgap ranged from ~4 to 1.35 eV. As $|B|$ increased, Im($E$) was enhanced, reflecting the amplification of non-Hermitian hopping by the Möbius correction. At 3% strain, the zigzag potential was approximately 0 meV when $B$ = 0. It was approximately 4 and 26 meV when $B$ = –0.75 and –1.5, respectively, and approximately 144 meV when $B$ = –3.

For both configurations at 3% strain, the gap was zero at $B$ = 0, and the Dirac point remained closed. At 6% strain for the armchair configuration, the K-path gap was ~687 meV at $B$ = 0. In fact, it increased monotonically with $|B|$ as follows: $B$ = −0.75 → ~24 meV, $B$ = −1.5 → ~50 meV, and $B$ = −3 → ~266 meV. At 12% strain for armchair, $B$ = 0 → ~8 meV and $B$ = −3 → ~492 meV. At 24% strain, the armchair gap of 1.35 eV at $B$ = −3 exceeded the zigzag gap of 0.78 eV at $B$ = −3. Armchair tension causes large Re-band separation more readily along the Γ–K–M path, whereas the zigzag response to $B$ is more complex. At high strain, the Re-bandgap is dominated by strain. At 24% strain with $B$ = 0, the zigzag and armchair gaps reached 54 and 0.98 eV, respectively. Notably, $B$ provides additional modulation: at $B$ = −3, the armchair and zigzag gaps reached 0.818 and 1.35 eV, respectively. Additionally, $B$ and strain synergistically enhanced the bandgap along the path.

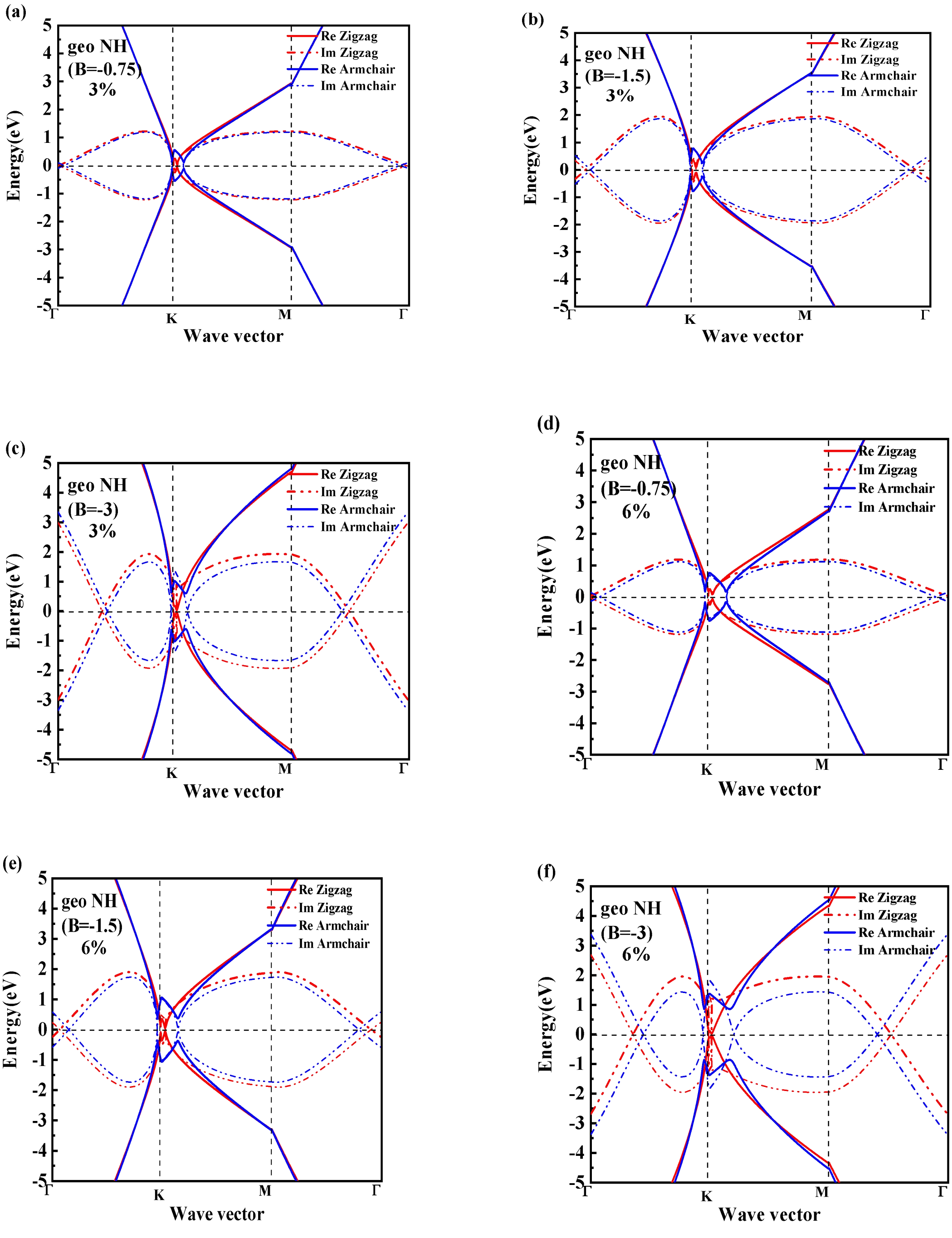

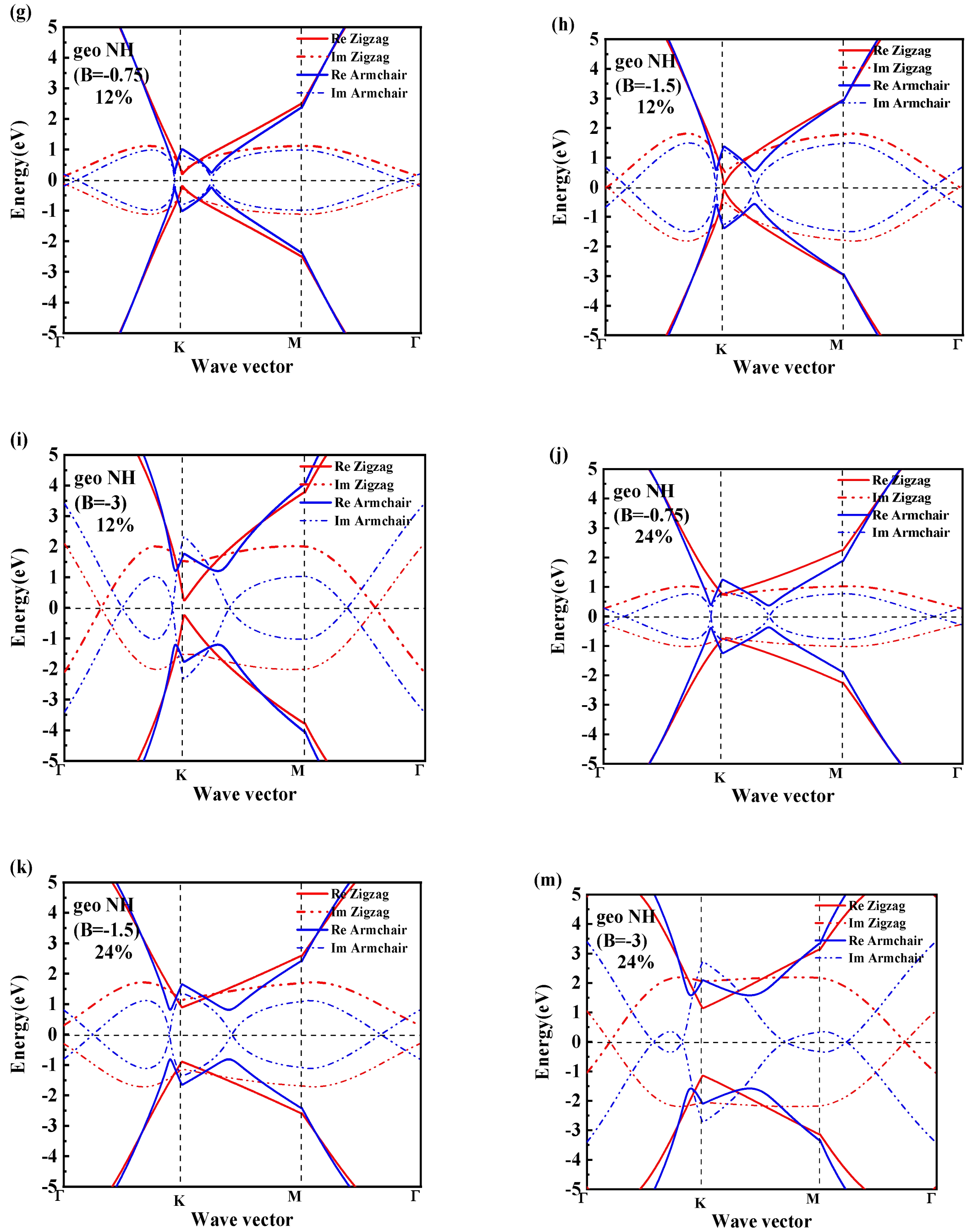

**Figure 6** Comparison of non-Hermitian band structure for zigzag and armchair edge configurations, showing real and imaginary components of energy as functions of strain $\varepsilon$ (3%, 6%, 12%, and 24%) and geometric parameter $B$ (0, −0.75, −1.5, and −3).

The band structure of the Tan–Bo geometric non-Hermitian model shows the electronic structure under zigzag and armchair tensile strains of 0%–24% and Möbius parameters $B$ = 0, −0.75, −1.5, and −3. Both the zigzag and armchair band structures were calculated using the same Tan–Bo two-band TB model; the only difference was whether strain was applied along the zigzag or armchair direction. The strain-dependent hopping employed the hybrid-Pereira distance scaling multiplied by the Möbius ($B$) angular correction. Subsequently, the non-Hermitian case was obtained by performing the full Hermitian TB assembly on the geo or hybrid model. Under zigzag and armchair tension, the geo model can reach a bandgap of 1.35 eV, with a max|Im($E$)| bandwidth of approximately 3.4 eV, as shown in **Figure 6**. The Re($E$) and Im($E$) bands simultaneously describe the electronic structure along the Γ–K–M path and are closely related to the position of the Dirac point relative to the Γ–K–M path and the deformation of the Re($E$) bands. Strain alters the three nearest-neighbor hopping terms and bond vectors through the Pereira bond-length scaling. The zigzag and armchair cases differed only in the direction of $\varepsilon$, while the form of the Hamiltonian remained unchanged.

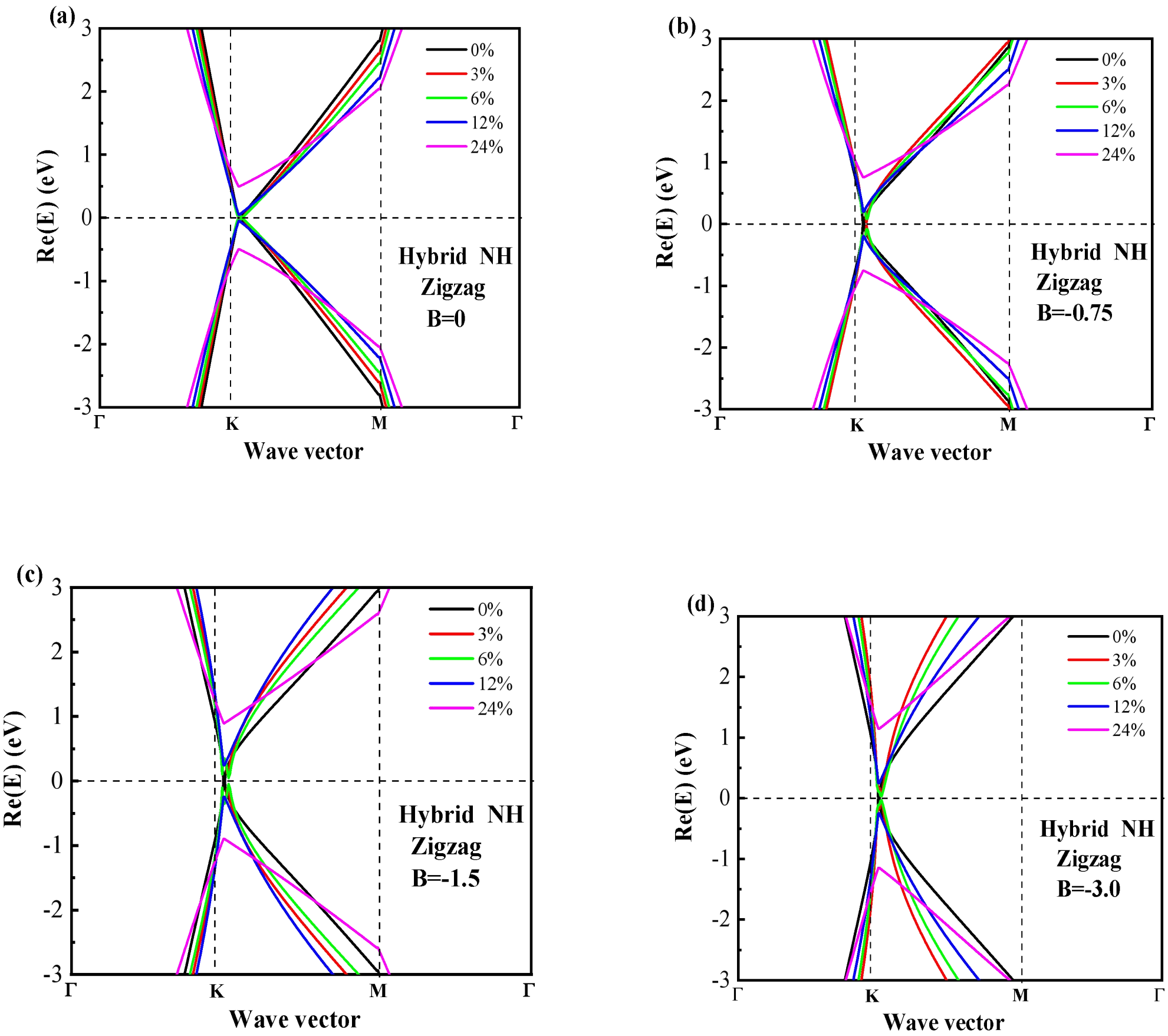

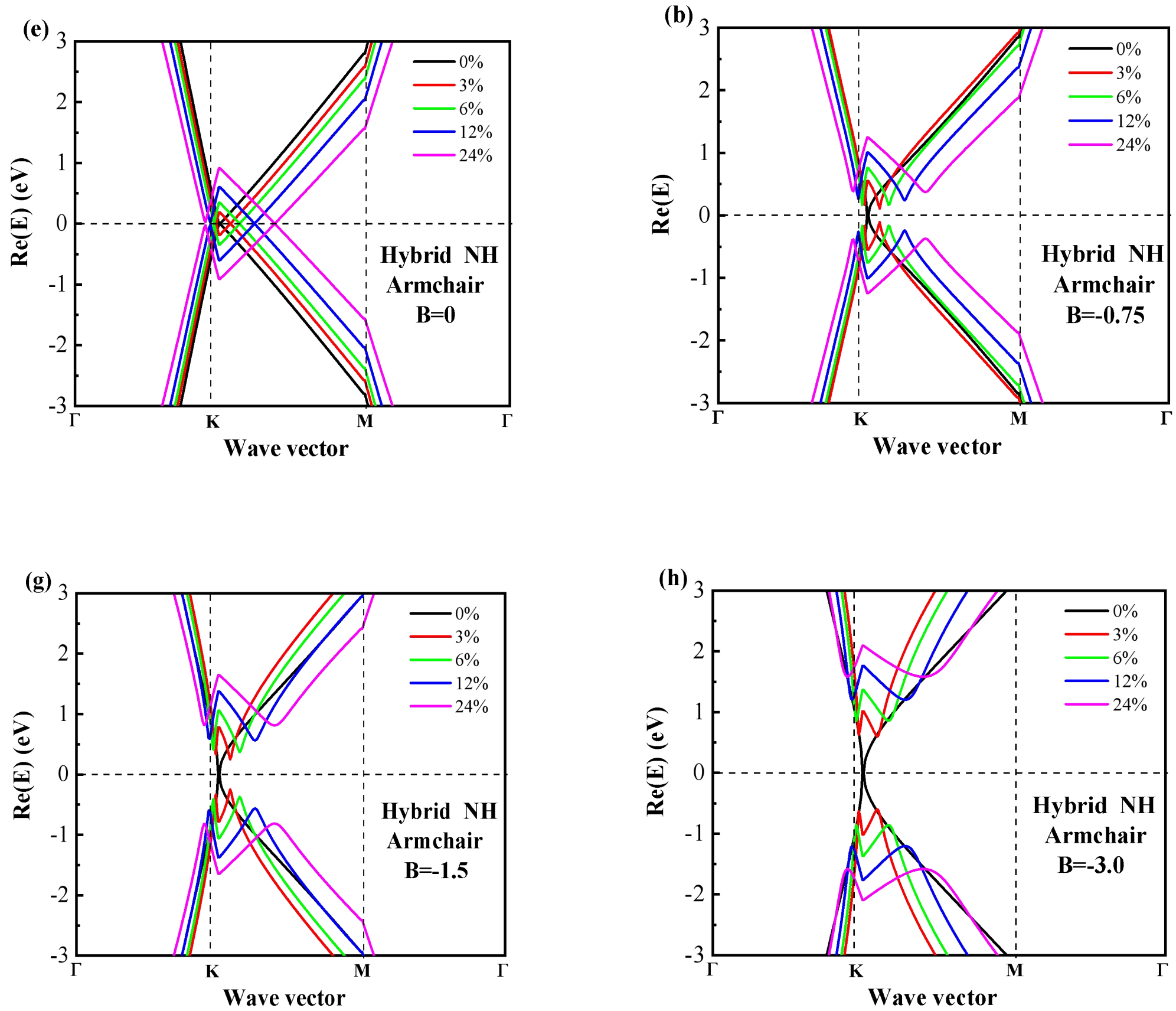

**Figure 7** Hybrid Tan–Bo non-Hermitian band structure showing real component of energy for zigzag and armchair edge configurations under different strains $\varepsilon$ (3%, 6%, 12%, and 24%) and geometric parameters $B$ (0, −0.75, −1.5, and −3).

Upon introducing $B \neq 0$, the hybrid Tan–Bo model developed a nonzero imaginary component and max|Re($E$)| increased with |$B$|, as shown in **Figure 7**. At 3% strain, the zigzag graphene exhibited a bandgap of 0 eV at $B$ = 0, and approximately 55, 222, and 74 meV at $B$ = -0.75, -1.5, and -3, respectively. At 24% strain, the gap was 0 eV at $B$ = -0.75 and -1.5, whereas it reached 2.416 eV at $B$ = -3 for the zigzag graphene, reflecting the amplification of non-Hermitian hopping by the Möbius correction. For both the zigzag and armchair configurations at $B$ = 0, the exact bandgap was 0 meV, and the Dirac point remained closed. At 6% armchair strain, the K-path gap was 8 meV at $B$ = 0, which increased with |$B$| to 334, 734, and 1716 meV at $B$ = -0.75, -1.5, and -3, respectively. At 12% armchair strain, it was 26 meV at $B$ = 0 and reached 2.606 eV at $B$ = -3. At 24% armchair strain with $B$ = -3, the bandgap was 3.164 eV, which exceeded the 2.416 eV gap for the zigzag graphene at $B$ = -3, indicating that armchair tension causes large Re-band separation more readily along the Γ–K–M path, whereas the zigzag response to $B$ is more

complex. In summary, strain dominates the Re-band deformation while $B$ provides additional modulation. Together, they synergistically enhance the bandgap along the path.

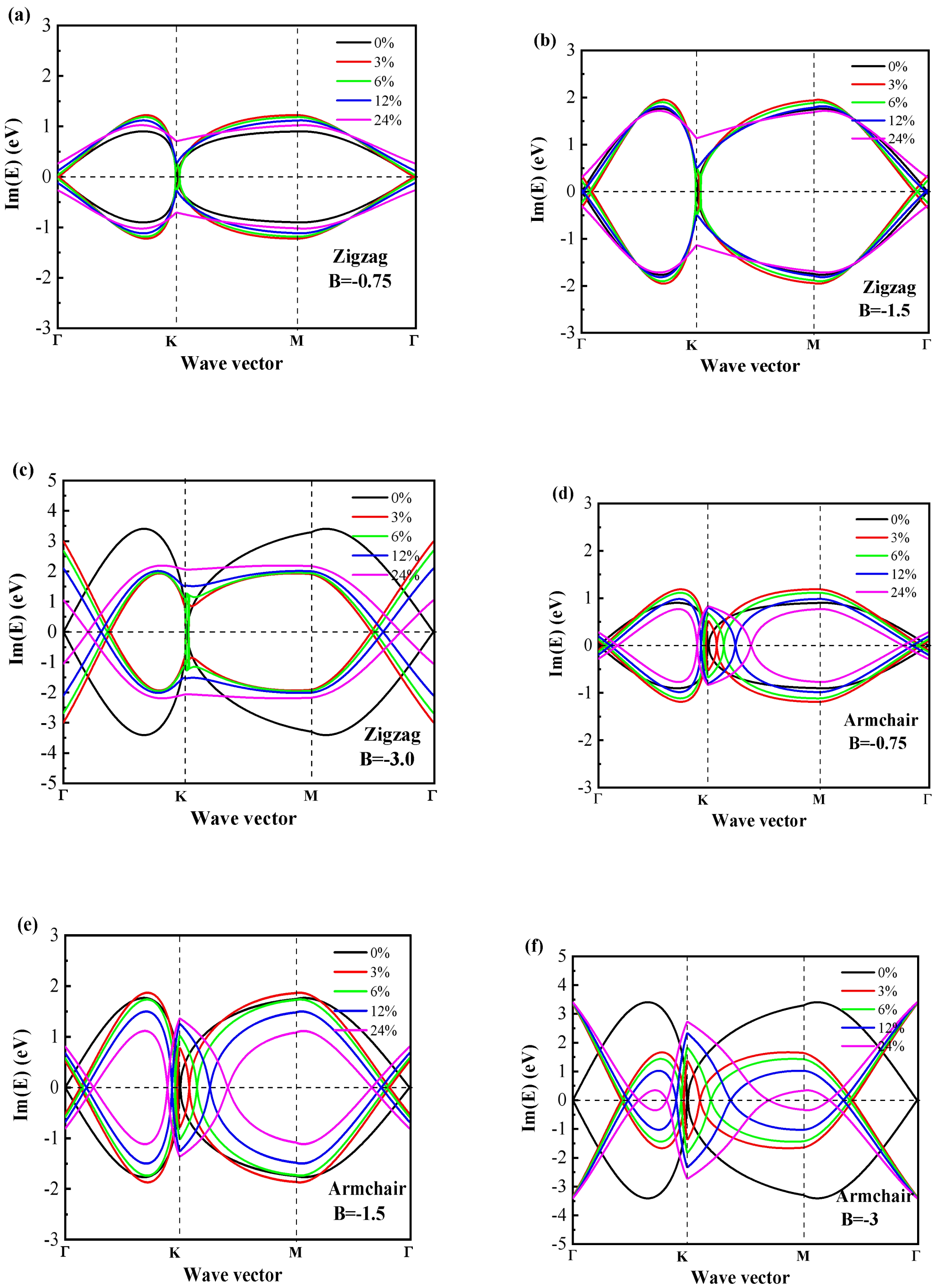

**Figure 8** Hybrid Tan–Bo non-Hermitian band structure showing imaginary component of energy for zigzag and armchair edge configurations under different strains $\varepsilon$ (3%, 6%, 12%, and 24%) and geometric parameters $B$ (0, -0.75, -1.5, and -3).

**Figures 8** present the dispersion of Im($E$) along the Γ–K–M–Γ path for the hybrid Tan–Bo model at $B$ = −0.75, −1.5, and −3. The Im lower and Im upper bands were approximately symmetric about Im = 0, approached zero near Γ, and formed a pair of "imaginary lobes" along the Γ–K and K–M segments. As $|B|$ increased, the Im amplitude along the entire path increased significantly. For the zigzag graphene, the bandwidth max|Im| increased from 0.9–1.2 eV at $B$ = −0.75 (**Figure 8a**) to1.7–2.0 eV at $B$ = -1.5 (**Figure 8b**) and reached ~3.4 eV at 0% strain for $B$ = -3 (**Figure 8c**). For the armchair configuration, the Im structure near K was more complex, with more pronounced crossings and narrowing displayed. Additionally, max|Im| was 0.9–1.2 eV at $B$ = -0.75 (**Figures 8d**), 1.4–1.9 eV at $B$ = −1.5(**Figures 8e**), and reached ~3.4 eV at 0% strain for $B$ = −3(**Figures 8f**), which is consistent with the geo limit. At the K-point, Im($E$) was small or exhibited a pinch feature. However, as $|B|$ and strain increased, the |Im| at K increased from 0 to 0.5–2.7 eV (at armchair $B$ = −3 and 24% strain, the |Im| at the K-point was ≈ 2.72 eV), indicating that the non-Hermitian imaginary component accumulates primarily along path segments outside the immediate vicinity of K instead of being uniformly distributed over the entire path. As strain increased, both the Re-path K gap and Im amplitude were synergistically modulated by strain and $B$, whereas the armchair and zigzag responses differed.

Physically, $B$ serves as a "knob" or control of the hybrid model: it introduces a Möbius angular correction on top of the Pereira strain-dependent hopping. $B$ = 0 yields a Hermitian baseline consistent with the experimental strain-gap scaling; increasing $|B|$ simultaneously amplifies both the Re-band separation and non-Hermitian imaginary component. At $B$ = −3, the hybrid model almost coincided with the geo model for the armchair configuration. Compared with experimentally reported exact bandgaps, reading the gap solely along the nominal K point of the Γ–K–M path would systematically overestimate the strain-induced gap, with the deviation increasing with $|B|$.

## 3.5 Optical conductivity under Tan–Bo anisotropy

To connect geometry-dependent hopping to a spectroscopically accessible observable, we evaluate the real interband optical conductivity $\sigma(\omega)$ of the Hermitian Tan–Bo Hamiltonian. Non-Hermitian eigenvalues are deliberately excluded: $\mathrm{Im}\ E$ arising from $t(-\mathrm{dr}) \neq t(\mathrm{dr})^*$ diagnoses nonreciprocal assembly and must not be interpreted as a quasiparticle lifetime in a closed-system Kubo response.

We use the Kubo–Greenwood formula on a dense rectangular k-mesh ($n_k = 96$ ) with Lorentzian broadening $\eta = 0.12$ eV, chemical potential $\mu = 0$, and velocity operators $v_\alpha = \partial H/\partial k_\alpha$. Spectra are reported in units of $\sigma_0$. The chosen $\eta$ smooths the $\delta$-function interband threshold for numerical integration: it is adequate for the mid-infrared universal plateau ($\hbar\omega \sim 0.8 - 1.5$eV) but does broaden the ultraviolet van Hove structure ($\sim 5.6$eV). Supplementary tests at $\eta = 0.05, 0.10$, and $0.15$ eV($B = 0, n_k = 96$) shift the van Hove maximum by $\lesssim$ 80 meV while leaving $\sigma_{xx}/\sigma_{yy}$ dichroism trends unchanged; the $\sim 1$ eV blueshift relative to the $\sim 4.6$eV excitonic resonance[26] therefore remains dominated by missing excitonic/self-energy corrections in independent-particle TB rather than by Lorentzian width alone. Residual fine ripples on $\sigma_\omega$(typically $\lesssim 0.05\sigma_0$ away from van Hove cusp regions at $n_k = 96$ ) reflect Brillouin-zone integration noise; they are far smaller than the anisotropy splittings ($\sigma_{xx}/\sigma_{yy}$ differences of order 0.1 – 0.3 in strained panels) that we emphasize.

**Figure 9** compares equilibrium isotropic ($B = 0$) and geometric ($B = -3$) Tan–Bo models. For $B = 0$, which coincides with standard nearest-neighbor TB / Slater–Koster at $r = r_0$, both $\sigma_{xx}$ and $\sigma_{yy}$ recover a mid-infrared response of order $\sigma_0$ and a $\pi - \pi^*$ van Hove peak near $\hbar\omega \approx 5.55\text{eV} \approx 2|t|$, which is consistent with $|E(M)| \simeq 2.8$eV based on comparison with experimental results (isotropic optics). Suspended and supported graphene both exhibit a nearly universal absorbance of (2.3 ± 0.2)% in the infrared–visible window, equivalent to the sheet conductivity $\sigma = \sigma_0 = \pi e^2/2h$ .[27] Our isotropic mid-infrared $\mathrm{Re}\,\sigma/\sigma_0$ is of order unity ( $\approx 0.81$ near 1.5eV at $n_k = 96$ ), matching the experimental scale; small residual deficits are attributable to finite mesh density, Lorentzian broadening, and the absence of doping/excitonic corrections in the independent-particle Kubo formula. Broadband measurements extending to the ultraviolet report a $\pi - \pi^*$ resonance near $\sim 4.6$eV after excitonic renormalization[26], whereas nearest-neighbor TB (and isotropic Tan–Bo) place the van Hove peak at $\sim 2|t| \approx 5.6$ eV. The computed peak shape therefore tracks the M-point joint density of states, while the absolute energy is systematically blueshifted relative to experiment — a known limitation of non-excitonic TB optics[28] Isotropic $B = 0$ also yields $\sigma_{xx} \approx \sigma_{yy}$, consistent with the absence of in-plane dichroism in unstrained graphene.

For $B = -3$, bond-resolved hoppings become strongly anisotropic. Spectral weight is redistributed: a low-energy feature appears in the visible window while the ultraviolet van Hove structure is reshaped in parallel with the Hermitian M-point upshift and the ratio $\sigma_{xx} \approx \sigma_{yy}$ develops a clear dichroism [**Fig. 9(c)**]. Physically, this visible feature arises from interband weight redistribution in a strongly anisotropic nearest-neighbor dispersion (effective $v_F$ and transition-matrix anisotropy), not from a new direct gap at $K$. In real samples it would compete with Drude tails, disorder broadening, and phonon-assisted backgrounds, and is unlikely to appear as an isolated resonance in chemical vapor deposition (CVD) or supported graphene. This result is consistent with our labelling it a pure single-particle model signature rather than an experimental line. Thus the Möbius parameter $B$ is readable in polarized optics even when the Hermitian Dirac point remains closed. We stress that there is no experimental "$B = -3$ graphene" sample: $B = -3$ is an illustrative anisotropy dial, not an optimum from $\mathcal{L}(B)$. The experimentally relevant counterpart is polarization dichroism under lattice anisotropy or uniaxial strain as discussed below rather than a direct validation of the Möbius B parameter itself.

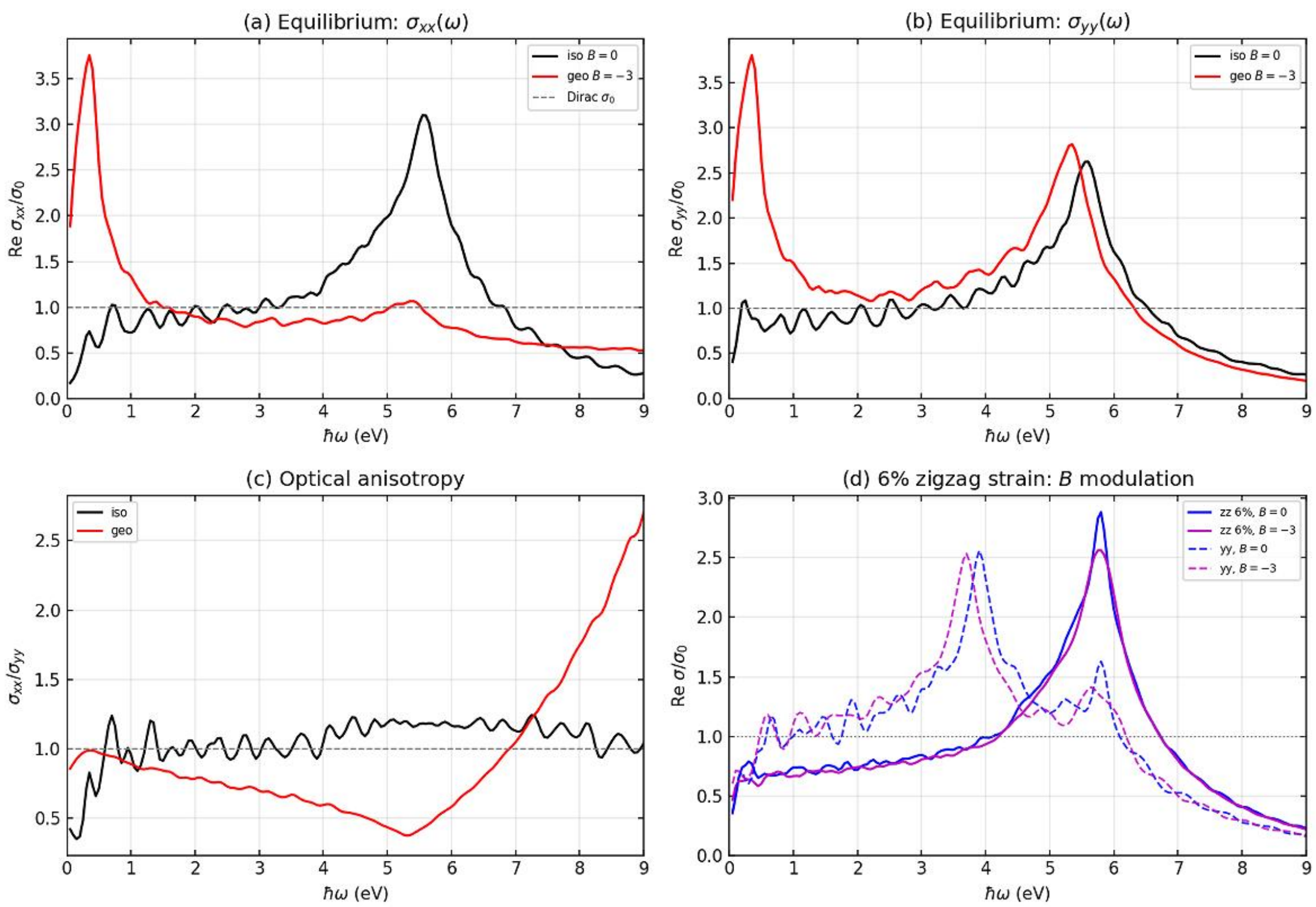

**Figure 9**: Equilibrium and strained optical conductivity of Hermitian Tan–Bo graphene (Kubo–Greenwood;$\sigma_0 = e^2/4\hbar = \pi e^2/2h$; $\eta = 0.12\text{eV}$; $n_k = 96$).(a–c) Equilibrium isotropic ($B = 0$) vs. illustrative geometric ($B =- 3$) presets: (a)$\text{Re}\sigma_{xx}/\sigma_0$; (b)$\text{Re}\sigma_{yy}/\sigma_0$; (c) anisotropy ratio $\sigma_{xx}/\sigma_{yy}$. (d) Zigzag 6% hybrid strain at $B = 0$ and $B =- 3$ (solid:$\sigma_{xx}$; dashed:$\sigma_{yy}$). Grey dashed line:$\sigma/\sigma_0 = 1$. Residual ripples ($\lesssim 0.05\sigma_0$) are integration noise. The visible feature at $B =- 3, \varepsilon = 0$ in (d) is a model interband-weight signature, not an experimental resonance (masked by disorder/Drude in real samples).

Under uniaxial zigzag tension, we use the hybrid hopping of $t_j(\varepsilon) = \frac{\phi(\omega_j)}{\phi_{eq}(\omega_j^0)} t_j^{Pereira/SK}(\varepsilon)$ so that Pereira/Harrison distance scaling sets the overall energy scale while B supplies angular modulation. **Figure 10** shows that increasing ε suppresses the geometric low-energy peak, blueshifts the $\sigma_{xx}$ van Hove maximum, and redshifts the $\sigma_{yy}$ peak, producing a strain-tunable optical dichroism. Finite $|B|$ further reshapes these peaks relative to $B = 0$.

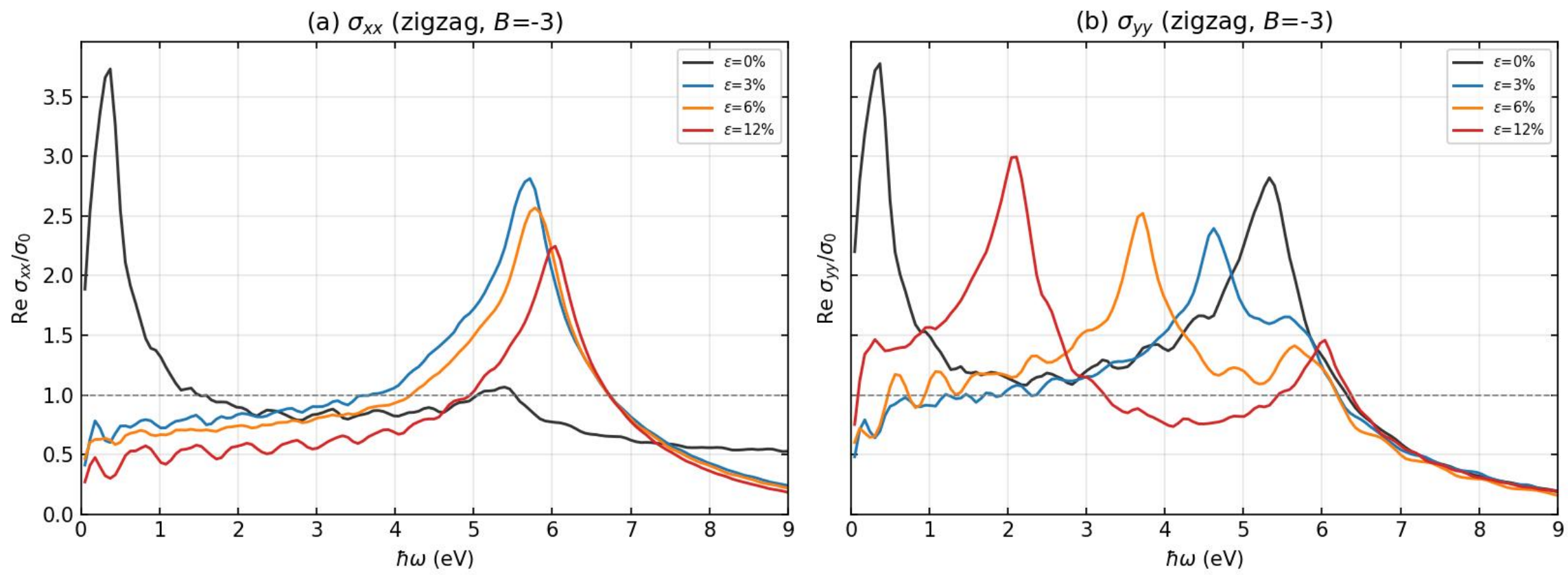

**Figure 10:** Strain-tuned optical conductivity for hybrid Tan–Bo with illustrative $B = -3$ under uniform zigzag tension ($\varepsilon = 0, 3\%, 6\%, 12\%$; $\varepsilon \gtrsim 6\%$ are extrapolations). (a) $\mathrm{Re}\,\sigma_{xx}/\sigma_0$; (b) $\mathrm{Re}\,\sigma_{yy}/\sigma_0$. Curves for increasing $\varepsilon$ use progressively darker shades (see plot legend); each panel compares polarization components at fixed strain. Hermitian assembly only; $\sigma_0 = \pi e^2/2h$; $\eta = 0.12\mathrm{eV}$; $n_k = 96$.

Comparison with experiment (strain optics). Uniaxial strain breaks $C_3$ and can induce intrinsic electronic optical anisotropy ($\sigma_{xx} \neq \sigma_{yy}$) in an ideal single-crystal sheet. Polarization-periodic transmittance modulations in large-area CVD graphene have been reported[29], but such samples are often polycrystalline, wrinkled, and pre-strained; the measured contrast may combine intrinsic interband dichroism with morphological/strain-field inhomogeneity, which our uniform zigzag TB model does not capture. Photoelastic Kubo analyses of ideal uniaxial graphene predict orientation-dependent van Hove selection rules [30]; we compare to that class of trend, not to a one-to-one mapping onto multigrain CVD microstructure. Our hybrid strain series **Fig. 10** reproduces the same qualitative trend: increasing $\varepsilon$ separates and reshapes the $\sigma_{xx}$ and $\sigma_{yy}$ peaks, in line with polarization-dependent transmittance under uniaxial strain without claiming a quantitative fit to a specific $B$ value or sample morphology. Raman G/2D splitting under controlled uniaxial tension confirms that few-percent strain is experimentally accessible and crystallographically oriented [31], although Raman does not directly measure the electronic gap. Reliable laboratory strains are typically $\lesssim 2-5\%$; scans up to $12\% - 24\%$ in **Figs. 10** are theoretical extrapolations, and experimental comparison should emphasize the few-percent dichroism trend. At moderate strain we find no abrupt “large optical absorption edge” in $\sigma(\omega)$: this refers to the absence of a strong interband threshold shift in the few-percent window, not to a claim about transport or ARPES gaps. Under zigzag tension the spectral gap $\Delta E@k_D$ remains $\sim 0$ for $\varepsilon \lesssim 18\% - 23\%$; the Pereira cone-merger threshold near $\sim 23\%$ is an indirect bulk gap (Dirac points at K and $K'$ merge at distinct momenta) distinct from a direct K-point opening[32] — consistent with Raman strain studies that report no clear electronic gap at accessible strains. The sharp low-energy peak at $\varepsilon = 0, B = -3$ is a geometric-anisotropy signature of the model and should not be identified with a measured experimental resonance.

## 4. Conclusions

In this study, we have established the Tan–Bo geometry-dependent hopping model within the nearest-neighbor TB framework for graphene $\pi$ electrons. Moreover, we have systematically compared the isotropic ($B$ = 0), geometric anisotropic ($B$ = −3), and SK(Harrison) and Pereira exponential distance scalings and distinguished between Hermitian/non-Hermitian Hamiltonian assemblies, the Bernal bilayer McCann model, and zigzag and armchair uniaxial strains. The main conclusions are summarized as follows.

**(1)** Equivalence of equilibrium structure to SK model. The isotropic Tan–Bo ($B$ = 0) was fully consistent with SK and standard TB models ($|t|$ = 2.8 eV, $B_{opt}$ = 0). The geometric preset ($B$ = −3) broadened $|t|$ to 3.31 eV and introduced complex hopping ($\max|t(-\boldsymbol{dr}) - t(\boldsymbol{dr})^*|$ = 3.66 eV). Nonetheless, the Hermitian Dirac cone at the K-point remained closed.

**(2)** Under a complete non-Hermitian assembly, the non-Hermitian band energy at point $M$ was approximately $E \approx \pm(3.12 \pm 3.28\, i)$ eV. When $B$ = 0, Im ≡ 0, Im(E) arose from the noncommutative TB assembly defined by $t(-\boldsymbol{dr}) \neq t(\boldsymbol{dr})^*$, reflecting the $k$-space distribution. This should not be equated to the quasiparticle lifetime or open-system dissipation.

**(3)** In the Bernal two-layer model with shared Tan–Bo layers, the function $f(\boldsymbol{k})$ yielded $\Delta K \approx 2\delta$ ($\delta$ = 50 meV → 0.10 eV), where the iso and geo models coincided. The main peak of the two-layer DOS was approximately 29% lower than that of the single layer. The interlayer contributions, i.e., $\gamma_1$ and $\delta$, dominated the electronic structure of the bilayer.

**(4)** Strain and hybrid hopping. Under strain, the hybrid scheme $t_j(\varepsilon) = \left[\dfrac{\phi(w_j)}{\phi_{eq}(w_j)}\right] \times t_{\text{distance}}$ was adopted: $t_{\text{distance}}$ followed either the Harrison power law or the Pereira exponential law. At $B = 0$, only the distance law applied, whereas at $B \neq 0$, the Möbius ratio superimposed the bond-direction anisotropy. For $\varepsilon = 3\%$, 6%, 12%, and 24% and $B = 0$, $-0.75$, $-1.5$, and $-3$, both the zigzag and armchair configurations exhibited synergistic modulation by the strain and $B$: increasing $|B|$ increased both the Re-band separation and non-Hermitian Im($E$).

(5) Hermitian Kubo $\sigma(\omega)$ recovers the experimental universal scale $\sigma_0$, places the TB van Hove peak above the ~4.6 eV excitonic UV resonance in shape, and yields strain dichroism qualitatively consistent with polarized transmittance trends in strained CVD graphene without assigning lifetime meaning to Im E or fitting a specific B.

## Acknowledgements

This study was financially supported by research grants from the Science and Technology Research Project of Fuling District, Chongqing (Grant Nos. FLKJ-2024BAG5128 and FLKJ2026AAG2033), the Science and Technology Research Program of the Chongqing Municipal Education Commission (Grant No. KJZD-K202501406), the Program of the Chongqing Municipal Education Commission (Grant No. KJQN-202401420), and the Science and Technology Research Project of Fuling District, Chongqing (Grant No. FLKJ-2024BAG5133).